# Bibliometric statistical properties of the 100 largest European universities: prevalent scaling rules in the science system

**Anthony F. J. van Raan**
Centre for Science and Technology Studies
Leiden University
Wassenaarseweg 52
P.O. Box 9555
2300 RB Leiden, the Netherlands

***Abstract***

*For the 100 largest European universities we studied the statistical properties of bibliometric indicators related to research performance, field citation density and journal impact. We find a size-dependent cumulative advantage for the impact of universities in terms of total number of citations. In previous work a similar scaling rule was found at the level of research groups. Therefore we conjecture that this scaling rule is a prevalent property of the science system. We observe that lower performance universities have a larger size-dependent cumulative advantage for receiving citations than top-performance universities. We also find that for the lower-performance universities the fraction of not-cited publications decreases considerably with size. Generally, the higher the average journal impact of the publications of a university, the lower the number of not-cited publications. We find that the average research performance does not 'dilute' with size. Evidently large universities, particularly top-performance universities are characterized by 'big and beautiful'. They succeed in keeping a high performance over a broad range of activities. This most probably is an indication of their overall scientific and intellectual attractive power. Next we find that particularly for the lower-performance universities the field citation density provides a strong cumulative advantage in citations per publication. The relation between number of citations and field citation density found in this study can be considered as a second basic scaling rule of the science system. Top-performance universities publish in journals with significantly higher journal impact as compared to the lower performance universities. We find a significant decrease of the fraction of self-citations with increasing research performance, average field citation density, and average journal impact.*

## 1. Introduction

In previous articles (van Raan 2006a, 2006b, 2007) we presented an empirical approach to the study of the statistical properties of bibliometric indicators of research groups. Now we focus on a two orders of magnitude larger aggregation level within the science system: the university. Our target group consists of the 100 largest European universities. We will distinguish between different 'dimensions': *top-* and *lower-performance* universities, *higher* and *lower field citation densities*, and *higher* and *lower journal impact*. In particular, we are interested in the phenomenon of size-dependent (size of a university in terms of number of publications) *cumulative advantage*[1] of impact

[1] By 'cumulative advantage' we mean that the dependent variable (for instance, number of citations of a university, ***C***) increases in a disproportional, nonlinear (in this case: power law) way as a function of the independent variable (for instance, in the present study the size of a research university, in terms of number of publications, ***P***). Thus, larger universities (in terms of ***P***) do not just receive more citations (as can be expected), but they do so increasingly more advantageously: universities that are twice as large as other universities receive, on average, about 2.5 more citations.

(in terms of numbers of citations), for different levels of research performance, field citation density and journal impact.

Katz (1999) discussed scaling relationships between number of citations and number of publications across research fields and countries. He concluded that the science system is characterized by cumulative advantage, more particularly a size-dependent 'Matthew effect' (Merton 1968, 1988). As explained in footnote 1, this implies a nonlinear increase of impact with increasing size, demonstrated by the finding that the number of citations as a function of number of publications (in Katz' study for 152 fields of science) exhibits a power law dependence with an exponent larger than 1. In our previous articles (van Raan 2006a, 2006b, 2007) we demonstrated a size-dependent cumulative advantage of the correlation between number of citations and number of publications also at the level of research groups. In this study we extent our observations to the level of entire universities.

We focus on performance-related differences of bibliometric properties of universities. Particularly important are the citation characteristics of the research fields in which a university is active (the field citation densities) and the impact level of the journals used by a university. Seglen (1992, 1994) found a poor correlation between the impact of publications and journal impact at *the level of individual publications*. However, grouping publications in classes of journal impact yielded a high correlation between publication and journal impact. This grouping is determined by journal impact classes, and not by a 'natural' grouping such as research groups and universities. In our previous study we showed a significant correlation between the average number of citations per publication of research groups, and the average journal impact of these groups. In this study we investigate whether this finding also holds at the level of entire universities.

The structure of this study is as follows. Within a set of the 100 largest universities in Europe we distinguish in our analysis between *performance*, *field citation densities* and *journal impact*. In Section 2 we discuss the data material of the universities, the application of the method, and the calculation of the indicators. In Section 3 we analyse the data of the 100 largest European universities in the framework of size-dependent cumulative advantage and classify the results of the analysis in main observations. Our analysis of performance- and field density-related differences of bibliometric properties of universities reveals further interesting results, particularly on the role of journal impact. These observations are discussed in the last part of Section 3. Finally, in Section 4 we summarize the main outcomes of this study.

## 2. Basic data and indicators derived from these data

We studied the statistics of bibliometric indicators on the basis of all publications (as far as published in journals covered by the Citation Index, 'CI publications'[2]) of the 100 largest European universities for the period 1997-2004[3]. This material is quite unique. To our knowledge no such compilations of very accurately collected publication sets on a large scale are used for statistical analysis of the characteristics of indicators at the university level. Obtaining data at the university level is not a trivial matter. The delineation of universities through externally available data such as the address information in the CI database is very problematic. For a thorough discussion of this problem, see Van Raan (2005a). The (CI-) publications were collected as part of a large

---

[2] Thomson Scientific, the former Institute for Scientific Information (ISI) in Philadelphia, is the producer and publisher of the Citation Index system covered by the Web of Science. Throughout this article we use the acronym CI (Citation Index) to refer to this data system.

[3] We included Israel. We have left out Lomonosov University of Moscow. As far as number of publications concerns, this university is one of the largest in Europe (about 24,000 publications in the covered 8-year period) but the impact is so low (***CPP/FCSm*** about 0.3) that it would have a very outlying position in the ranking.

EC study on the scientific strengths of the European Union and its member states[4]. For a detailed discussion of methodological and technical issues we refer to Moed (2006). From a listing of more than 250 European universities we selected the 100 largest. The period covered is 1997-2004 for both publications and citations received by these publications. In total, the analysis involves the work of many thousands of senior researchers in 100 large universities and covers around 1,5 million publications and 11 million citations (excluding self-citations), about 15% of the worldwide scientific output and impact.

The indicators are calculated on the basis of a total time-period analysis. This means that publications are counted for the entire period (1997-2004) and citations are counted up to and including 2004 (e.g., for publications from 1997, citations are counted in the period 1997-2004, and for publications from 2004, citations are counted only in 2004). We are currently updating our data system with the 2005 and 2006 publication and citation data.

We apply the CWTS[5] standard bibliometric indicators. Only 'external' citations, i.e., citations corrected for self-citations, are taken into account. An overview of these indicators is given in the text box here below. For a detailed discussion we refer to Van Raan (1996, 2004, 2005b).

**Standard Bibliometric Indicators:**

- Number of publications ***P*** in CI-covered journals of a university in the specified period;
- Number of citations ***C*** received by ***P*** during the specified period, without self-citations; including self-citations: ***Ci,*** i.e., number of self-citations ***Sc = Ci – C***, relative amount of self-citations ***Sc/Ci***;
- Average number of citations per publication, without self-citations (***CPP***);
- Percentage of publications not cited (in the specified period) ***Pnc***;
- Journal-based worldwide average impact as an international reference level for a university (***JCS***, journal citation score, which is our journal impact indicator), without self-citations (on this world-wide scale!); in the case of more than one journal we use the average ***JCSm***; for the calculation of ***JCSm*** the same publication and citation counting procedure, time windows, and article types are used as in the case of ***CPP***;
- Field-based[6] worldwide average impact as an international reference level for a university (***FCS***, field citation score), without self-citations (on this world-wide scale!); in the case of more than one field (as almost always) we use the average ***FCSm***; for the calculation of ***FCSm*** the same publication and citation counting procedure, time windows, and article types are used as in the case of ***CPP***; we refer in this article to the ***FCSm*** indicator as the 'field citation density';
- Comparison of the ***CPP*** of a university with the world-wide average based on ***JCSm*** as a standard, without self-citations, indicator ***CPP/JCSm***;
- Comparison of the ***CPP*** of a university with the world-wide average based on ***FCSm*** as a standard, without self-citations, indicator ***CPP/FCSm***;
- Ratio ***JCSm/FCSm*** is the relative, field-normalized journal impact indicator.

In Table 1 we show as an example the results of our bibliometric analysis for the first 30 universities within the European 100 largest. This table makes clear that our indicator calculations allow an extensive statistical analysis of these indicators for our set of universities. Of the above indicators, we regard the internationally standardized (field-normalized) impact indicator ***CPP/FCSm*** as our 'crown' indicator. This indicator enables us to observe immediately whether the performance of a university is significantly far below (indicator value < 0.5), below (0.5 - 0.8), around (0.8 - 1.2), above (1.2 – 1.5), or far above (>1.5) the international (Western world dominated) impact standard averaged over all fields (van Raan 2004).

---

[4] The ASSIST (*Analysis and Studies of Statistics and Indicators on Science and Technology*) project.

[5] Centre for Science and Technology Studies, Leiden University.

[6] We here use the definition of fields based on a classification of scientific journals into *categories* developed by Thomson Scientific/ISI. Although this classification is not perfect, it provides a clear and 'fixed' consistent field definition suitable for automated procedures within our data-system.

***Table 1**: Largest 30 European universities*

| | University | | ***P*** | *C* | *CPP* | *Pnc* | *CPP/ FCSm* |
|---|---|---|---|---|---|---|---|
| *1* | UNIV CAMBRIDGE | UK | ***36.349*** | 361.681 | 9,95 | 29,1 | 1,63 |
| *2* | UNIV COLL LONDON | UK | ***34.407*** | 346.028 | 10,06 | 26,9 | 1,46 |
| *3* | UNIV OXFORD | UK | ***33.780*** | 355.856 | 10,53 | 29,5 | 1,67 |
| *4* | IMPERIAL COLL LONDON | UK | ***27.017*** | 222.713 | 8,24 | 30,7 | 1,45 |
| *5* | LUDWIG MAXIMILIANS UNIV MUNCHEN | DE | ***23.519*** | 177.317 | 7,54 | 30,8 | 1,14 |
| *6* | UNIV PARIS VI PIERRE & MARIE CURIE | FR | ***23.468*** | 146.483 | 6,24 | 32,8 | 1,09 |
| *7* | UNIV MILANO | IT | ***23.006*** | 175.181 | 7,61 | 30,0 | 1,11 |
| *8* | UNIV UTRECHT | NL | ***22.668*** | 189.671 | 8,37 | 28,3 | 1,37 |
| *9* | KATHOLIEKE UNIV LEUVEN | BE | ***22.521*** | 153.851 | 6,83 | 34,9 | 1,22 |
| *10* | UNIV MANCHESTER | UK | ***22.470*** | 137.812 | 6,13 | 34,4 | 1,16 |
| *11* | UNIV WIEN | AT | ***21.940*** | 137.251 | 6,26 | 32,9 | 1,01 |
| *12* | UNIV ROMA SAPIENZA | IT | ***21.778*** | 119.076 | 5,47 | 37,7 | 0,95 |
| *13* | TEL AVIV UNIV | IL | ***21.447*** | 112.337 | 5,24 | 35,9 | 0,94 |
| *14* | UNIV HELSINKI | FI | ***21.034*** | 179.662 | 8,54 | 28,5 | 1,38 |
| *15* | LUNDS UNIV | SE | ***20.631*** | 157.944 | 7,66 | 27,9 | 1,21 |
| *16* | KAROLINSKA INST STOCKHOLM | SE | ***20.525*** | 213.629 | 10,41 | 23,2 | 1,30 |
| *17* | KOBENHAVNS UNIV | DK | ***19.555*** | 153.583 | 7,85 | 27,4 | 1,18 |
| *18* | UNIV AMSTERDAM | NL | ***19.333*** | 163.417 | 8,45 | 28,9 | 1,35 |
| *19* | UPPSALA UNIV | SE | ***18.998*** | 140.518 | 7,40 | 28,6 | 1,17 |
| *20* | RUPRECHT KARLS UNIV HEIDELBERG | DE | ***18.735*** | 155.451 | 8,30 | 30,1 | 1,22 |
| *21* | ETH ZURICH | CH | ***18.611*** | 148.078 | 7,96 | 29,8 | 1,52 |
| *22* | KINGS COLL UNIV LONDON | UK | ***18.601*** | 161.460 | 8,68 | 28,7 | 1,32 |
| *23* | HEBREW UNIV JERUSALEM | IL | ***18.389*** | 127.263 | 6,92 | 33,2 | 1,16 |
| *24* | UNIV PARIS XI SUD | FR | ***18.183*** | 115.157 | 6,33 | 32,8 | 1,13 |
| *25* | UNIV EDINBURGH | UK | ***17.786*** | 164.380 | 9,24 | 29,7 | 1,48 |
| *26* | HUMBOLDT UNIV BERLIN | DE | ***17.780*** | 127.381 | 7,16 | 31,6 | 1,13 |
| *27* | LEIDEN UNIV | NL | ***16.832*** | 147.821 | 8,78 | 26,9 | 1,26 |
| *28* | UNIV ZURICH | CH | ***16.783*** | 154.154 | 9,19 | 29,2 | 1,33 |
| *29* | UNIV BARCELONA | ES | ***16.783*** | 103.628 | 6,17 | 32,4 | 1,03 |
| *30* | UNIV BRISTOL | UK | ***16.387*** | 119.960 | 7,32 | 29,7 | 1,31 |

## 3. Results and Discussion

### 3.1 Impact scaling and research performance

In our previous study (van Raan 2006a, 2006b, 2007) we showed how a set of research groups is characterized in terms of the correlation between size (the total number of publications ***P*** of a specific research group[7]) and the total number of citations ***C*** received by a group. Now we calculated the same correlation for all 100 largest European universities. Fig. 3.1.1 shows that this correlation is described with a strong significance (coefficient of determination of the fitted regression is $R^2$ = 0.79) by a power law:

$$\boldsymbol{C}(\boldsymbol{P}) = 0.36\ \boldsymbol{P}^{1.31}\ .$$

At the lower side of ***P*** (and ***C***) we observe a few ‘outliers’. These are universities with a considerably lower number of citations as compared to the other larger universities (among them Charles University of Prague and the University of Athens). We observe that the size of universities leads to a cumulative advantage (with exponent $\boldsymbol{\alpha}=+1.31$) for the number of citations received by these universities. Thus, the Matthew effect also works in at the aggregation level of entire universities. The intriguing question is how the

[7] The number of publications is a measure of size in the statistical context described in this article. It is, however, a proxy for the real size of a research group or a university, for instance in terms number of *staff full time equivalents* (fte) available for research.

*research performance* of the universities (measured by the indicator ***CPP/FCSm***) relates to size-dependency. Gradual differentiation between top- and lower-performance (top/bottom 10%, 25%, and 50% of the ***CPP/FCSm*** distribution) enables us to study the correlation of ***C*** with ***P*** and possible scale effects (size-dependent cumulative advantage) in more detail. The results are presented in Figs. 3.1.2 - 3.1.4 and a summary of the findings in Table 3.

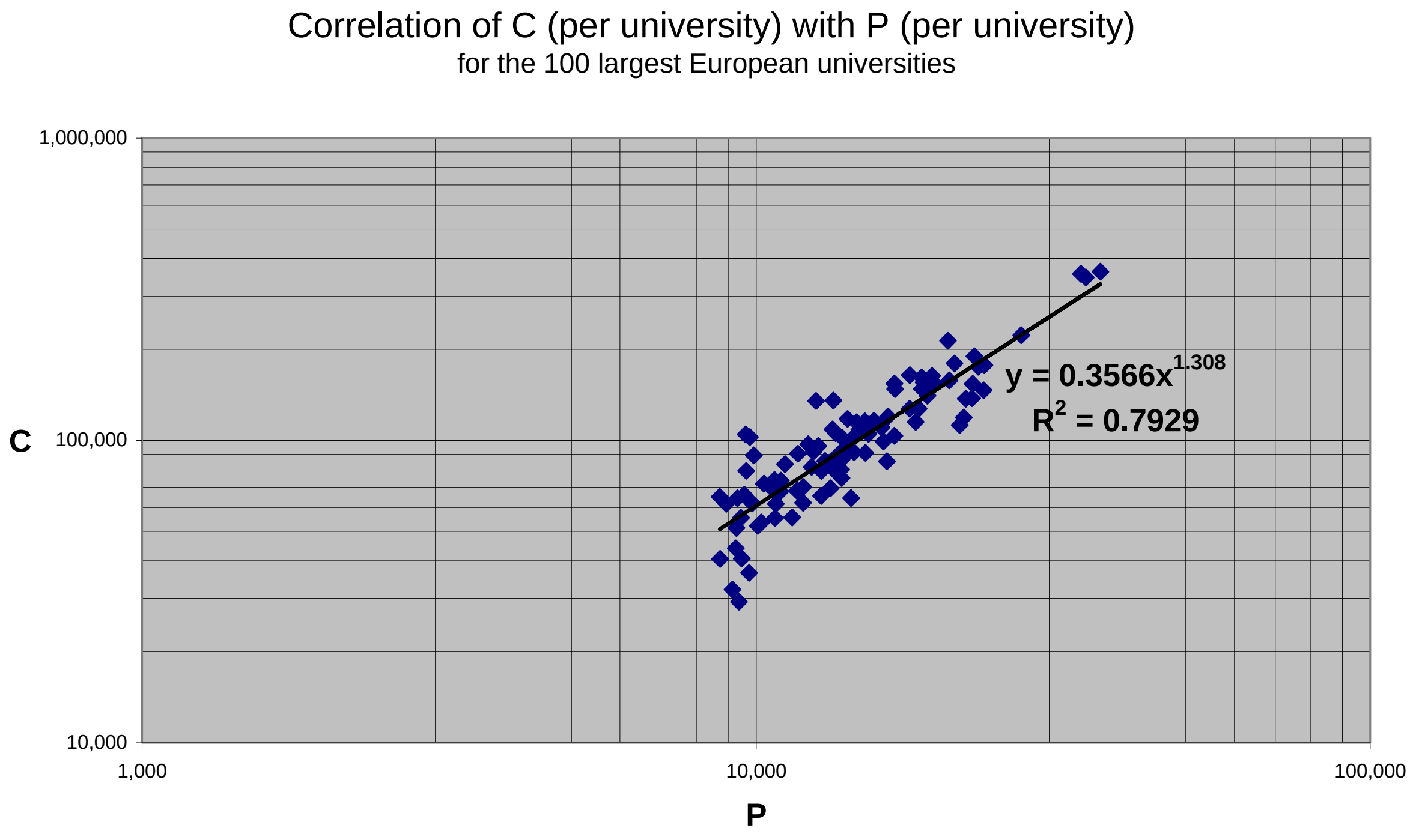

***Fig. 3.1.1***: *Correlation of the number of citations (**C**) received per university with the number of publications (**P**) of these universities for all 100 largest European universities.*

The group of highest performance universities (top-10%) does not have a cumulative advantage (i.e., exponent significantly[8] > 1). The bottom-10% exponent is heavily determined by the outliers. The broader top-25% shows a slight ($\alpha$=+1.16) and the bottom-25% a stronger cumulative advantage ($\alpha$=+1.33). If we divide the entire set of universities in a top- and bottom-50% we see that both subsets have more or less equal exponents. Thus, the most intriguing finding is that the *lowest* performance universities have a *larger* size-dependent cumulative advantage than top-performance universities. This phenomenon was already observed at the level of research groups (van Raan 2006a, 2006b, 2007). It is fascinating that within the science system this scaling rule covers at least two orders of magnitude in size of entities. Furthermore, the *top*-performance universities are generally the *larger* ones, i.e., in the right hand side of the correlation function.

[8] To estimate the influence of these noisy data, we randomly removed five universities. We found that the error in the exponent $\alpha$ is about ± 0.05. Thus, the noisiness of data remains within acceptable limits and does not substantially affect our findings.

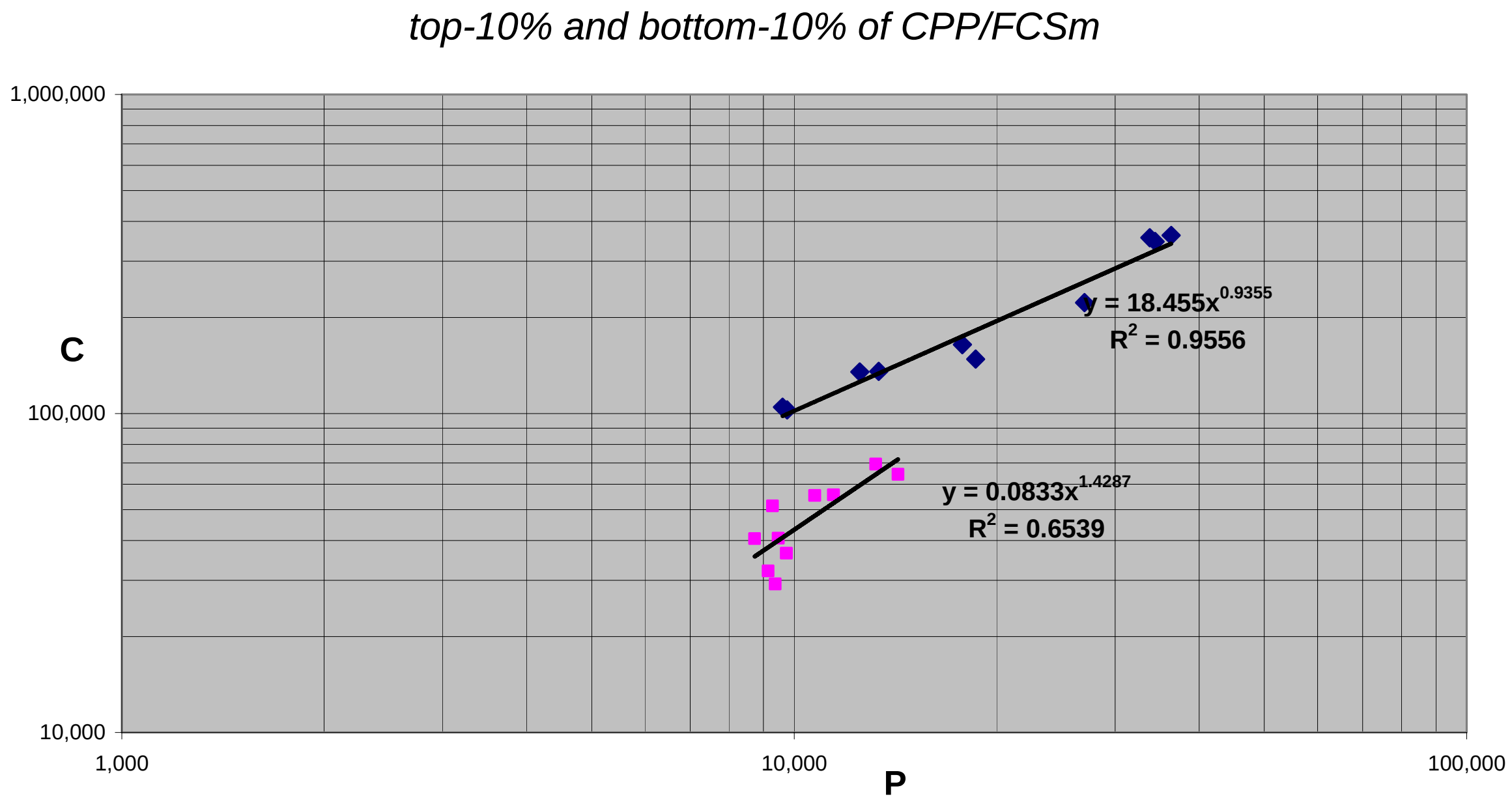

***Fig. 3.1.2:*** *Correlation of the number of citations (**C**) received per university with the number of publications (**P**) for the top-10% (of **CPP/FCSm**) universities (diamonds) and the bottom-10% universities (squares) within the 100 largest European universities.*

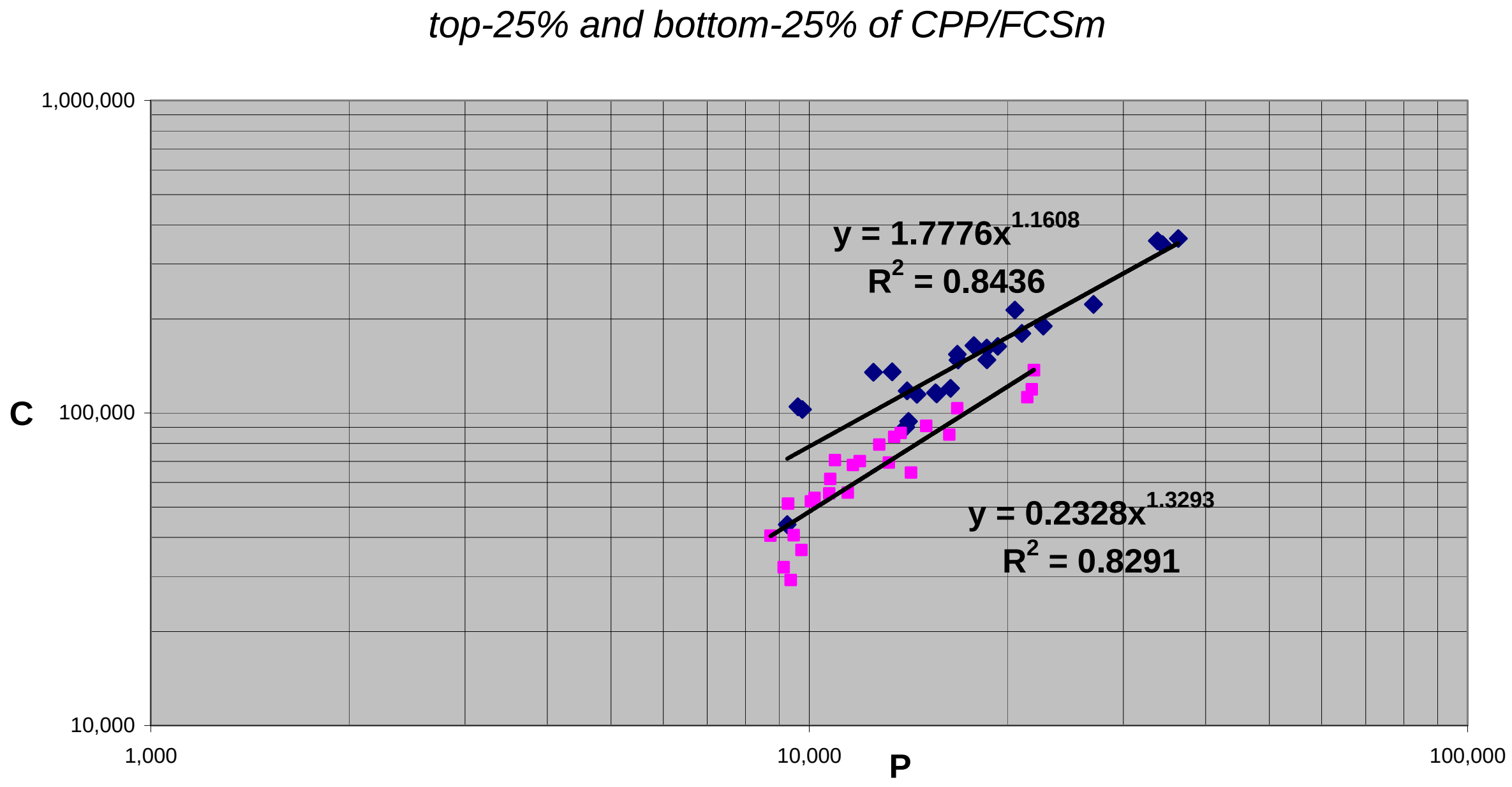

***Fig. 3.1.3:*** *Correlation of the number of citations (**C**) received per university with the number of publications (**P**) for the top-25% (of **CPP/FCSm**) universities (diamonds) and the bottom-25% universities (squares) within the 100 largest European universities.*

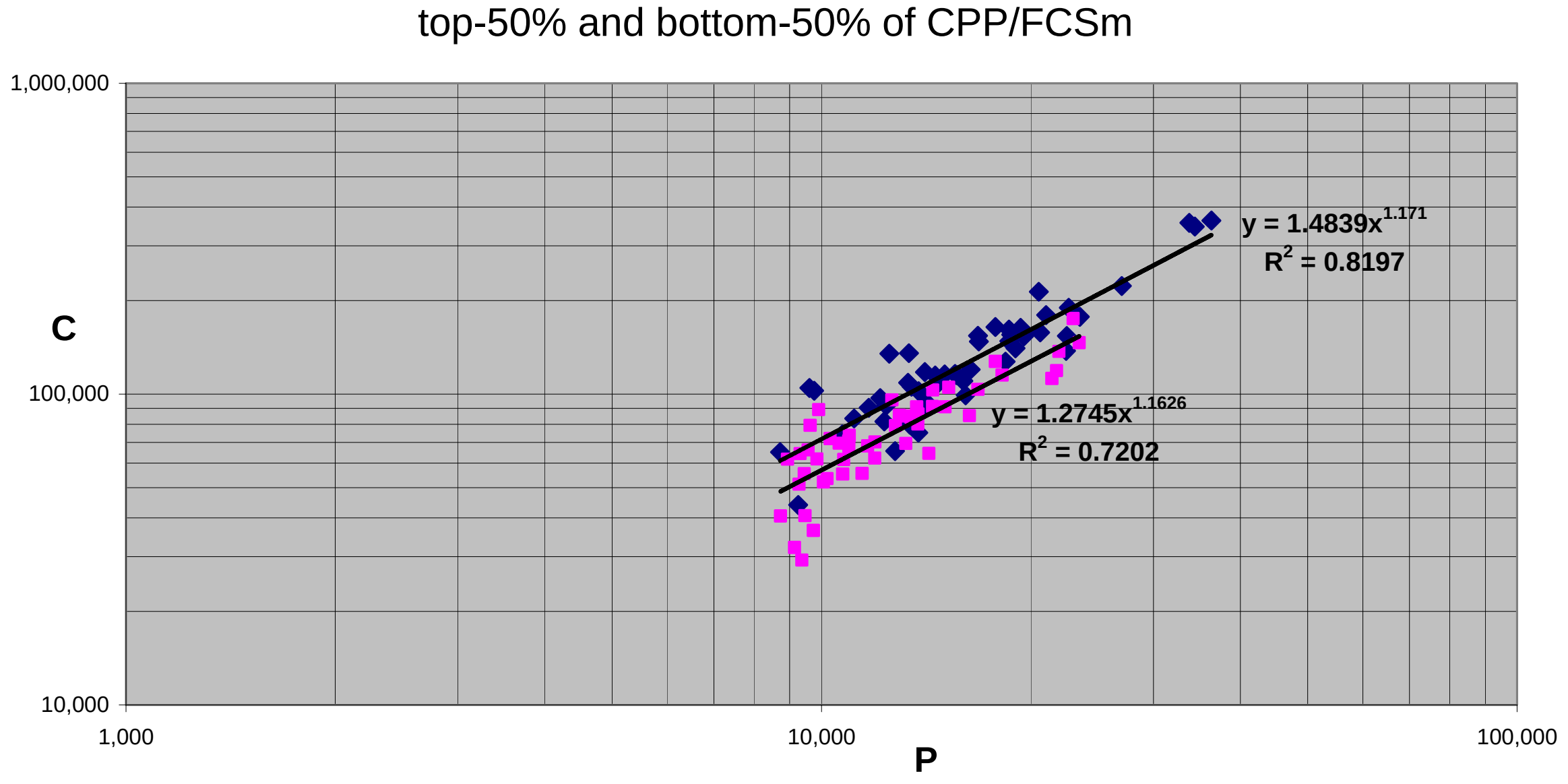

***Fig. 3.1.4:*** *Correlation of the number of citations (**C**) received per university with the number of publications (**P**) for the top-50% (of **CPP/FCSm**) universities (diamonds) and the bottom-50% universities (squares) with the 100 largest European universities.*

***Table 3.1**: Power law exponent **α** of the correlation of **C** with **P** for the 100 largest European universities in the indicated modalities. The differences in **α** between top and bottom modalities are indicated by **Δα**(b,t).*

| | |
|---|---|
| **All 100** | 1.31 |
| **top 10%** | 0.94 |
| **bottom 10%** | 1.43 |
| ***Δα**(b,t)* | *0.49* |
| | |
| **top 25%** | 1.16 |
| **bottom 25%** | 1.33 |
| ***Δα**(b,t)* | *0.17* |
| | |
| **top 50%** | 1.17 |
| **bottom 50%** | 1.16 |
| ***Δα**(b,t)* | *-0.01* |

An important feature of research impact is the number of *not-cited* publications. We analysed the correlation of the fraction (percentage) of not-cited-publications ***Pnc*** of the 100 largest European universities with size (***P***) of a university. The results are shown in Fig. 3.1.5. We observe that the fraction of not-cited publications decreases with low significance as a function of size. The significance of the correlation is too low for clear results. Thus, as a further step we investigate this correlation with a distinction between top- and lower-performance universities. Fig. 3.1.6 shows the results for the top- and bottom-25%, and Fig. 3.1.7 for the top-50% and bottom-50% of the ***CPP/FCSm*** distribution of the 100 largest universities.

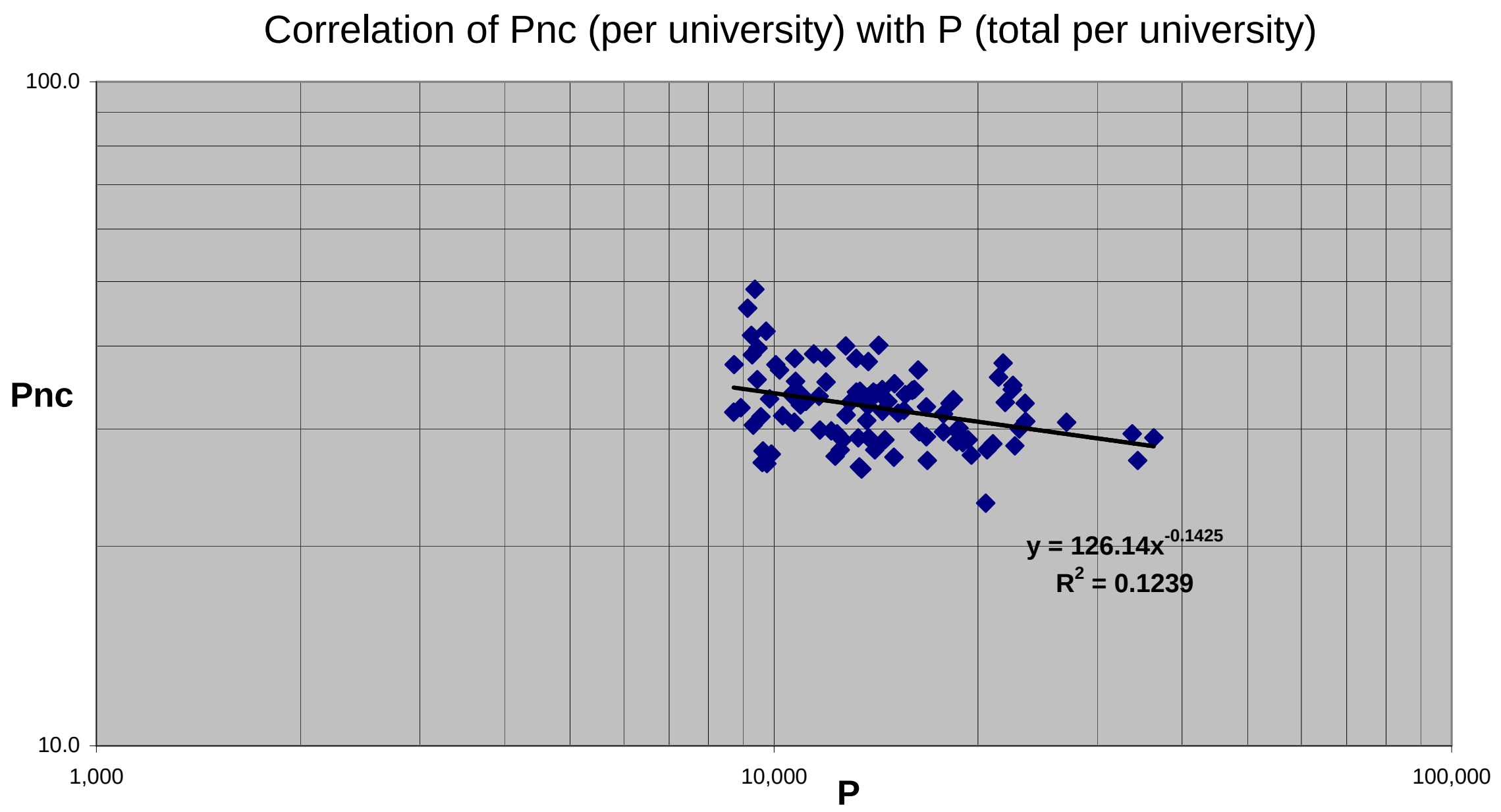

***Fig. 3.1.5:*** *Correlation of the percentage of not cited publications (**Pnc**) with the number of publications (**P**) for the entire set of the 100 largest European universities.*

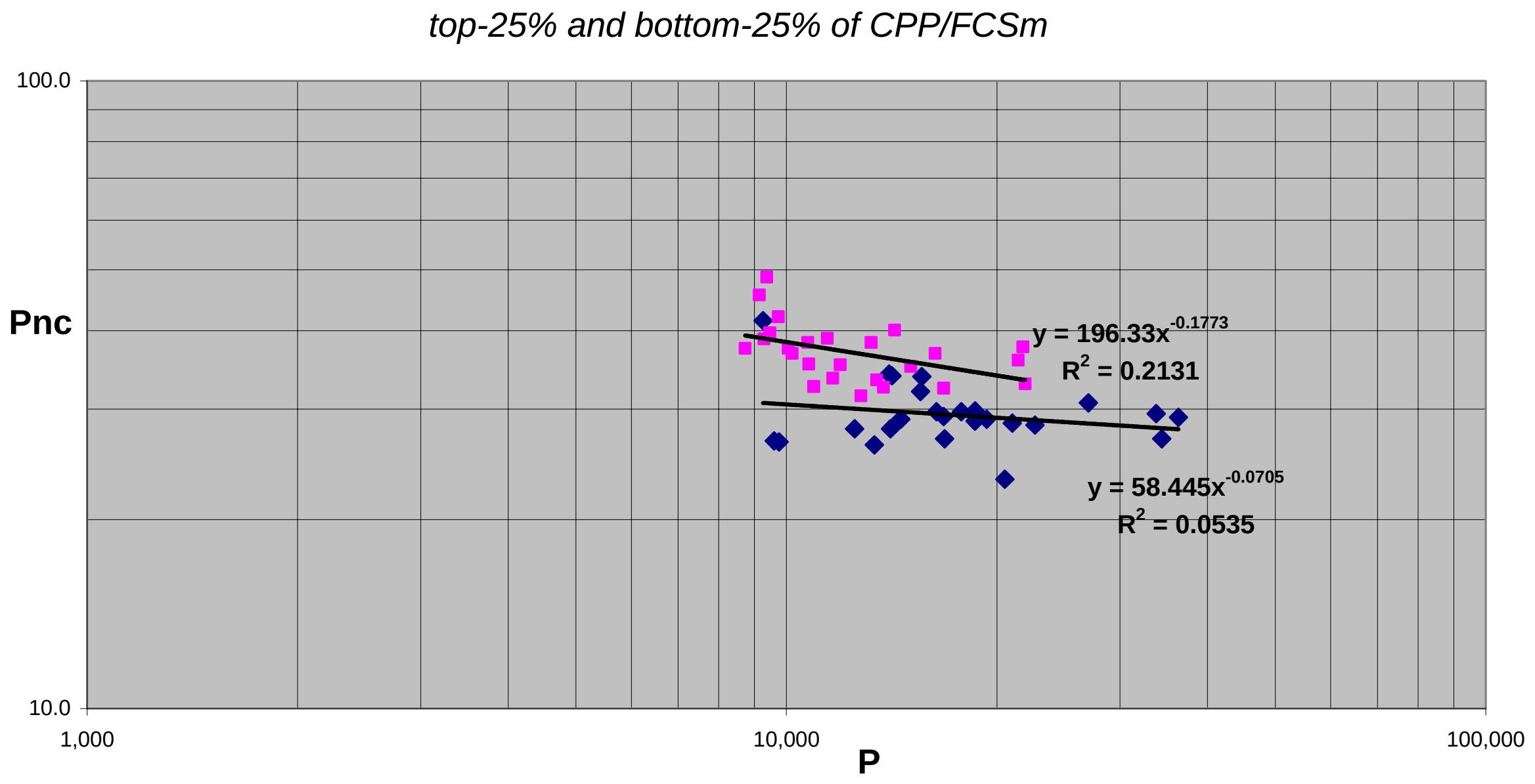

***Fig. 3.1.6:*** *Correlation of the relative number of not cited publications (**Pnc**) with the number of publications (**P**) for the top-25% (of **CPP/FCSm**) universities (diamonds), and the bottom-25% universities (squares).*

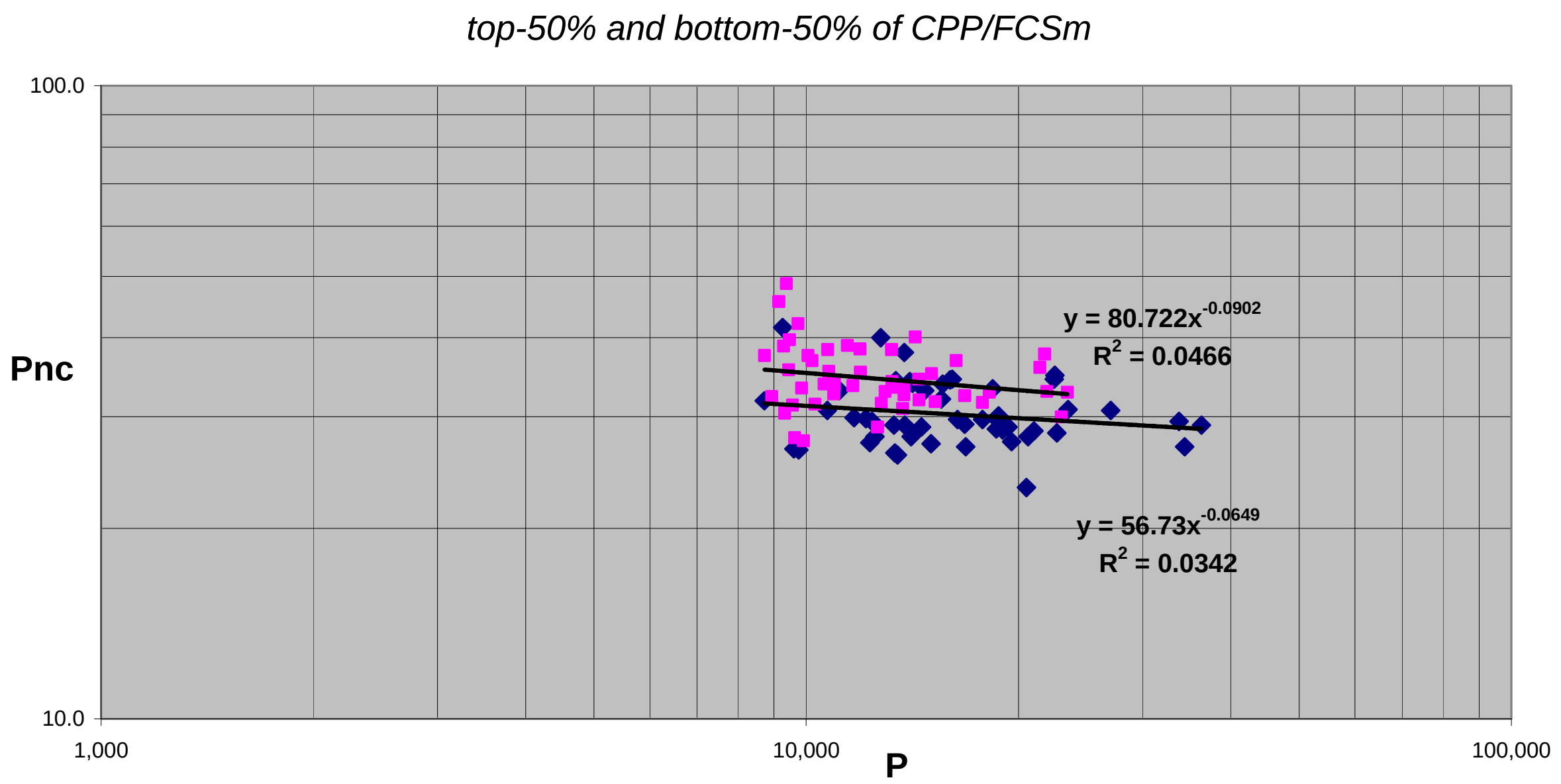

***Fig. 3.1.7:*** *Correlation of the relative number of not cited publications (**Pnc**) with the number of publications (**P**) for the top-50% (of **CPP/FCSm**) universities (diamonds), and for the bottom-50% universities (squares).*

The observations suggest that the fraction of non-cited publications decreases with size, particularly for the lower performance universities. This phenomenon was also found at the level of research groups (van Raan 2006a, 2006b, 2007) which means that we discovered another scaling rule in the science system covering at least two orders of magnitude. We notice, however, that this scaling rule for non-cited publications is less strong at the level of entire universities as compared to groups. Advantage by size works by a mechanism in which the number of not-cited publications is diminished. This mechanism works at the level of research groups as follows. The larger the number of publications in a group, the more those publications are 'promoted' which otherwise would have remained uncited. Thus, size reinforces an internal promotion mechanism, namely initial citation of these 'stay behind' publications in other more cited publications of the group. Then authors in other groups are stimulated to take notice of these stay behind publications and eventually decide to cite them. Consequently, the mechanism starts with within-group citation (which is not necessarily the same as self-citation), and subsequently spreads. It is obvious that particularly the lower performance groups will benefit from this mechanism. Top-performance groups do not 'need' the internal promotion mechanism to the same extent as low performance groups. This explains, at least in a qualitative sense, why top-performance groups show less, or even no cumulative advantage by size. Since an entire university is the sum of a large number of research groups, the above mechanism will also be visible at the university level.

We also investigated the relation between research performance as measured by indicator ***CPP/FCSm*** with size in terms of ***P***. We find a very slight positive correlation as shown in Fig. 3.1.8 for all 100 universities and in Fig. 3.1.9 for the top- and bottom-25% of the ***CPP/FCSm*** distribution. This, however, this is certainly not a cumulative advantage; the exponent of the correlation is very small, around 0.2. Probably the most interesting aspect of this measurement is that performance does not decrease, not 'dilute' with increasing size.

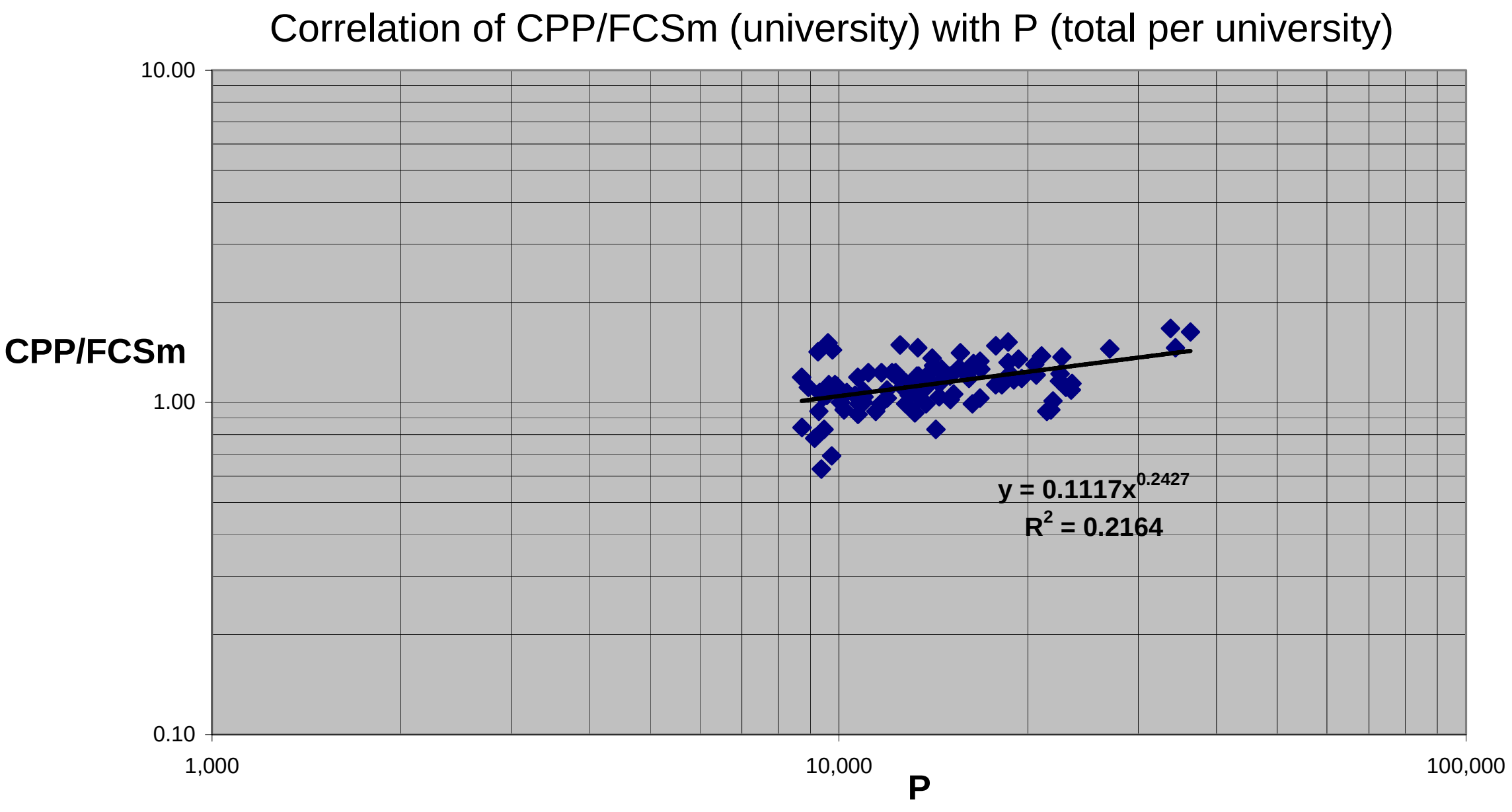

***Fig. 3.1.8:*** *Correlation of **CPP/FCSm** with the number of publications (**P**) for the entire set of all 100 largest European universities.*

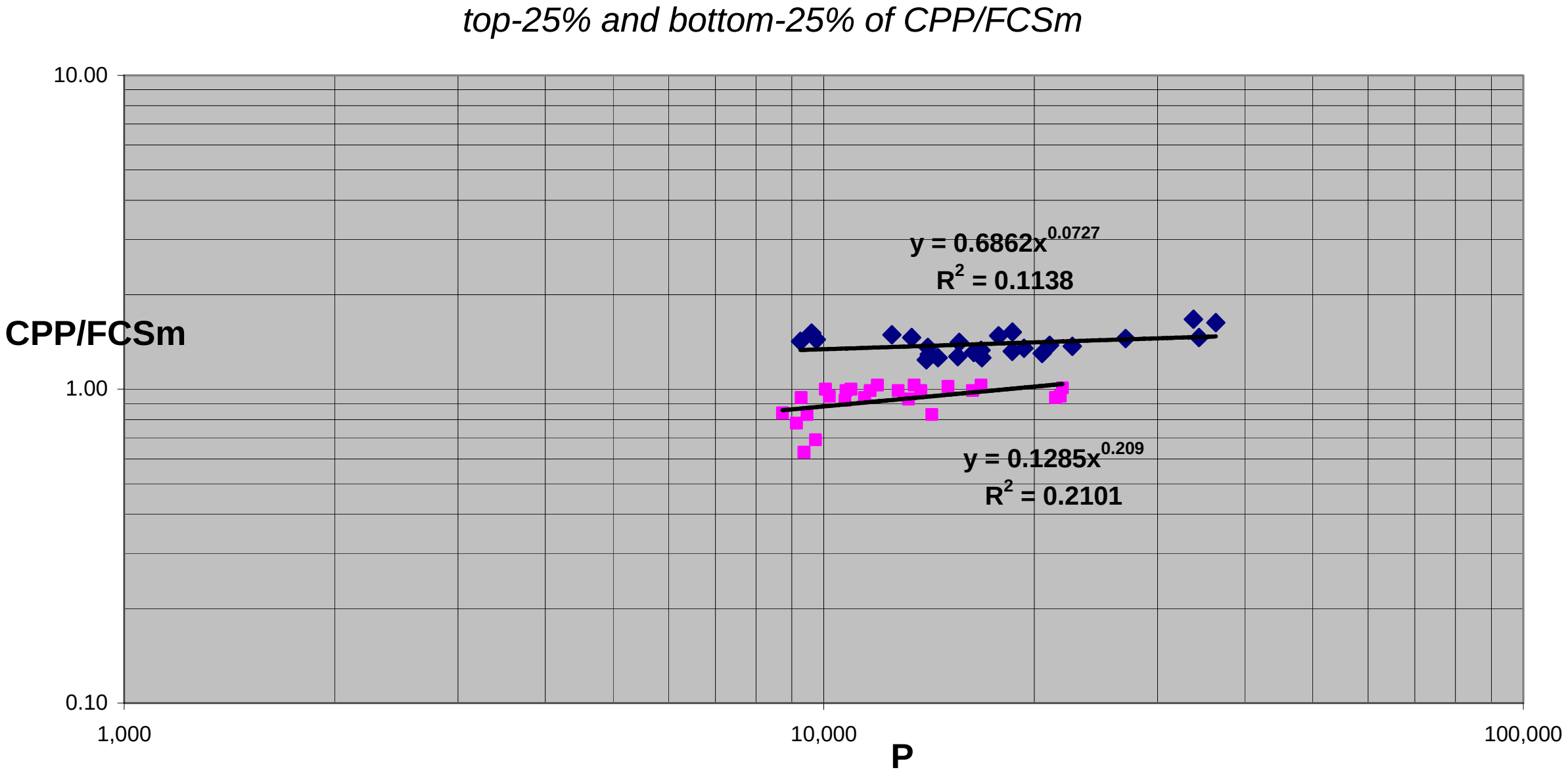

***Fig. 3.1.9:*** *Correlation of **CPP/FCSm** with the number of publications (**P**) for the top-25% (diamonds) and the bottom-25% (squares) of **CPP/FCSm** distribution of the 100 largest European universities.*

## 3.2 Impact scaling, field citation density and journal impact

In Fig. 3.2.1 we present the correlation of the number of citations with size for those universities among the 100 largest European universities that have *high* and *low* field citation densities, i.e., top-25% and bottom-25%, respectively, of the ***FCSm*** distribution. We observe that the high field density universities hardly have a cumulative advantage (exponent $\alpha$ = 1.09). The low field citation density universities have a considerably size-dependent cumulative advantage (exponent $\alpha$ = 1.50).

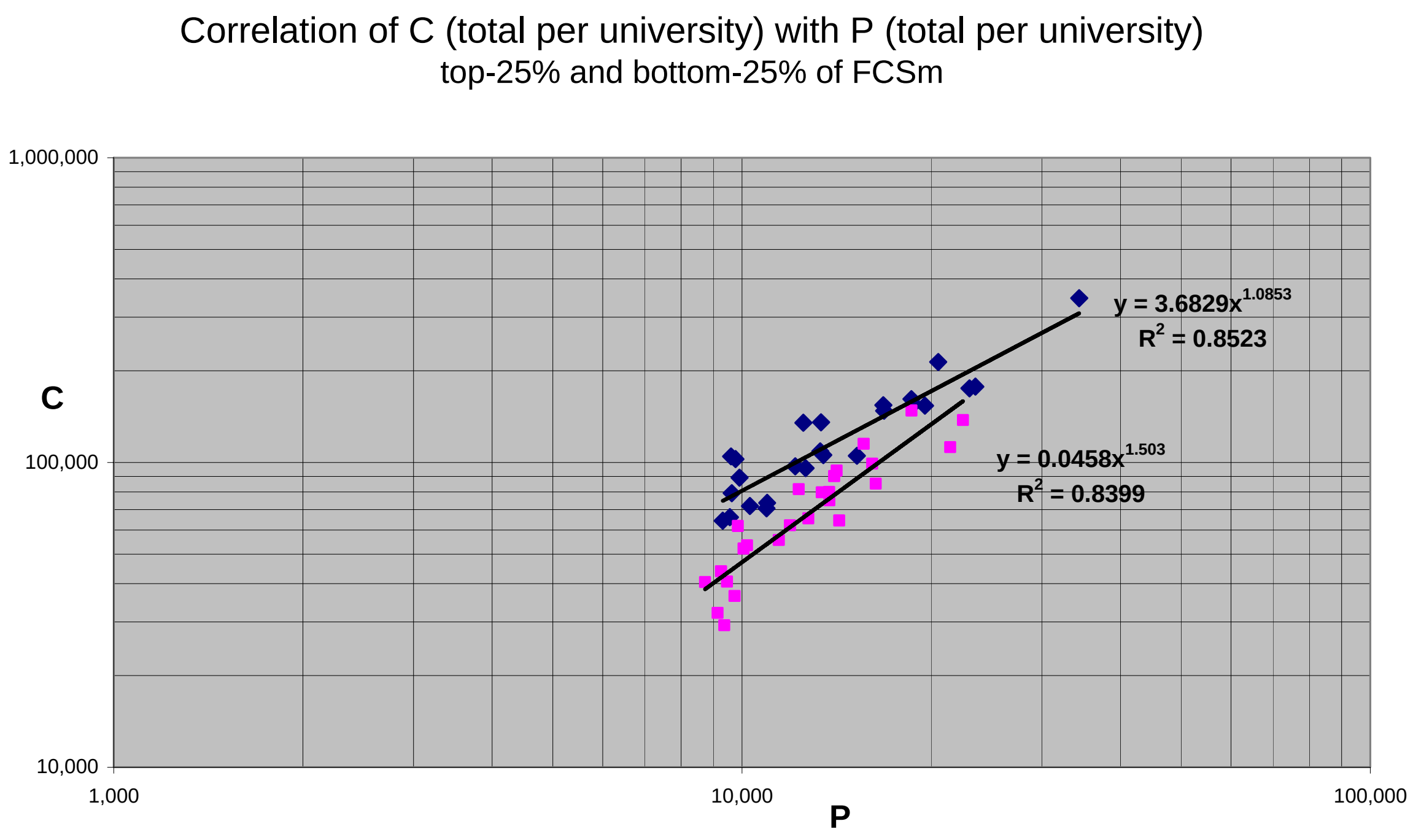

***Figure 3.2.1:*** *Correlation of the number of citations (**C**) with the number of publications (**P**) for the universities within the top- (diamonds) and the bottom-25% (squares) of the field citation density (**FCSm**) distribution.*

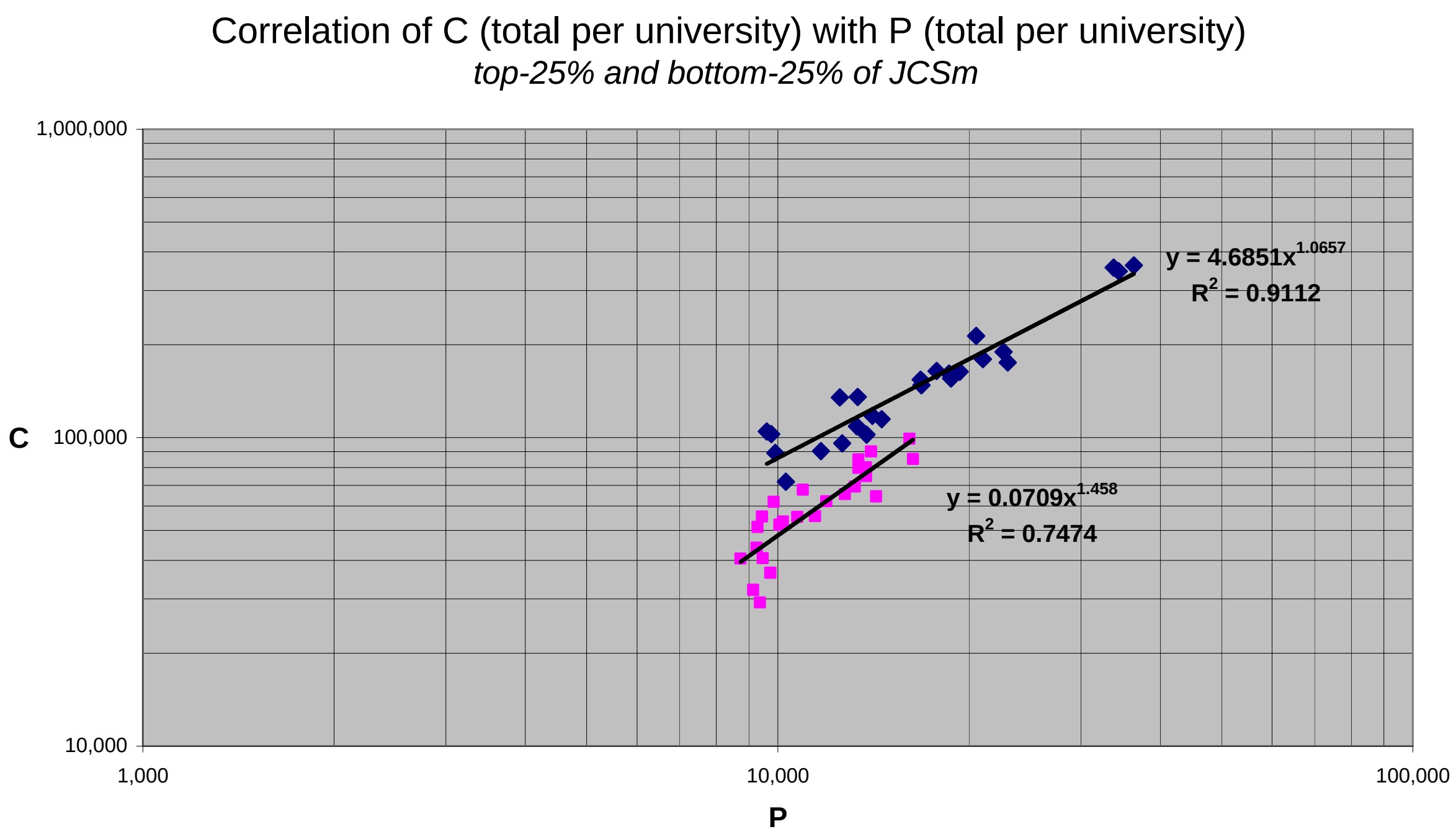

***Figure 3.2.2**: Correlation of the number of citations (**C**) with the number of publications (**P**) for the universities within the top- (diamonds) and the bottom-25% (squares) of the field citation density (**JCSm**) distribution.*

In Fig. 3.2.2 we present a similar correlation for the top- and bottom-25% of the ***JCSm***, the average journal impact of a university. We see that these results are practically the same as in Fig. 3.2.1. Given the strong correlation of ***JCSm*** and ***FCSm*** at the level of universities, as illustrated in Fig. 3.2.3, this similarity can be expected. We remark, however, that the correlation of ***JCSm*** and ***FCSm*** has a power exponent 1.22 which means that the ***JCSm*** values increase in a nonlinear way ('cumulatively') with ***FCSm***.

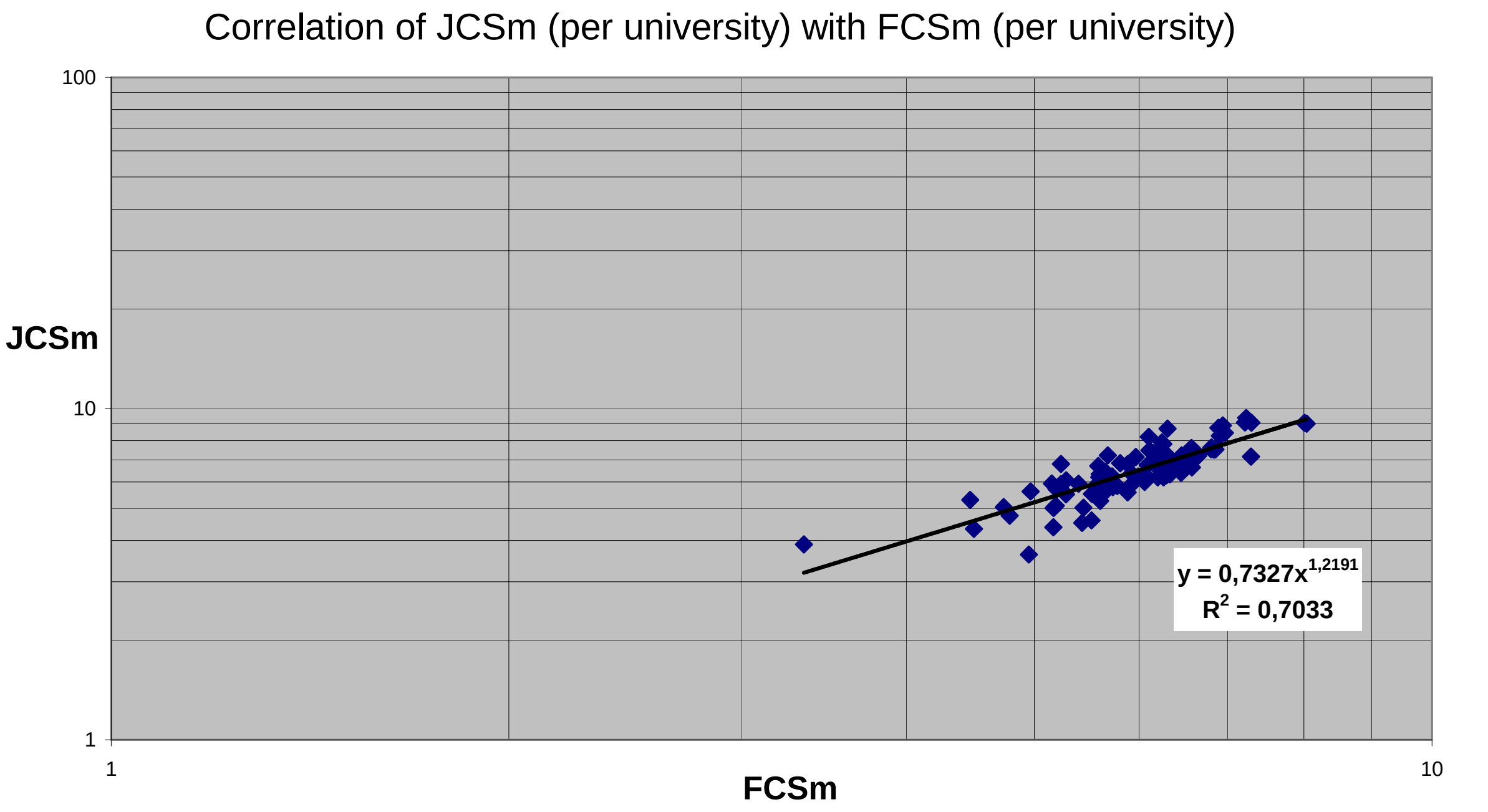

***Figure 3.2.3***: *Correlation of the average journal impact (**JCSm**) with the average field citation density (**FCSm**) for all 100 largest European universities.*

We now investigate the relation between citation impact of a university in terms of average number of citations per publication (***CPP***) on the one hand, and field citation density (***FCSm***) and journal impact (***JCSm***) on the other. Seglen (1994) showed that the citedness of *individual publications* ***CPP*** is not significantly affected by journal impact[9]. However, grouping publications in classes of journal impact yielded a high correlation between publication citedness and journal impact. We found that also a 'natural' grouping of publications, such as the work of a research group, leads to a high correlation of ***CPP*** and ***JCSm*** (van Raan 2006b, 2007).

In this study we find that this is also the case at the aggregation level of entire universities. We find a significant correlation between the average number of citations per publication for the 100 largest European universities (***CPP***), and both the field citation density (***FCSm***) as well as the average journal impact of these universities (***JCSm***). We applied again the distinction between top- and lower-performance universities in order to find performance-related aspects in the above relation. The results are shown for the correlation of ***CPP*** with ***FCSm*** for the entire set of all 100 largest European universities in Fig. 3.2.4, and for the top-performance (top-25% of ***CPP/FCSm***) and lower performance (bottom-25% of ***CPP/FCSm***) universities in Fig. 3.2.5. The correlation of ***CPP*** with ***JCSm*** for the entire set of all 100 largest European universities is presented in Fig. 3.2.6 and for the top-performance and lower performance universities in Fig. 3.2.7. We see hat these correlations are very significant.

[9] In Seglen's work journal impact was defined with the ISI (Web of Science) journal impact factor; he did not consider the more sophisticated journal impact indicators such as the ***JCSm*** used in this study.

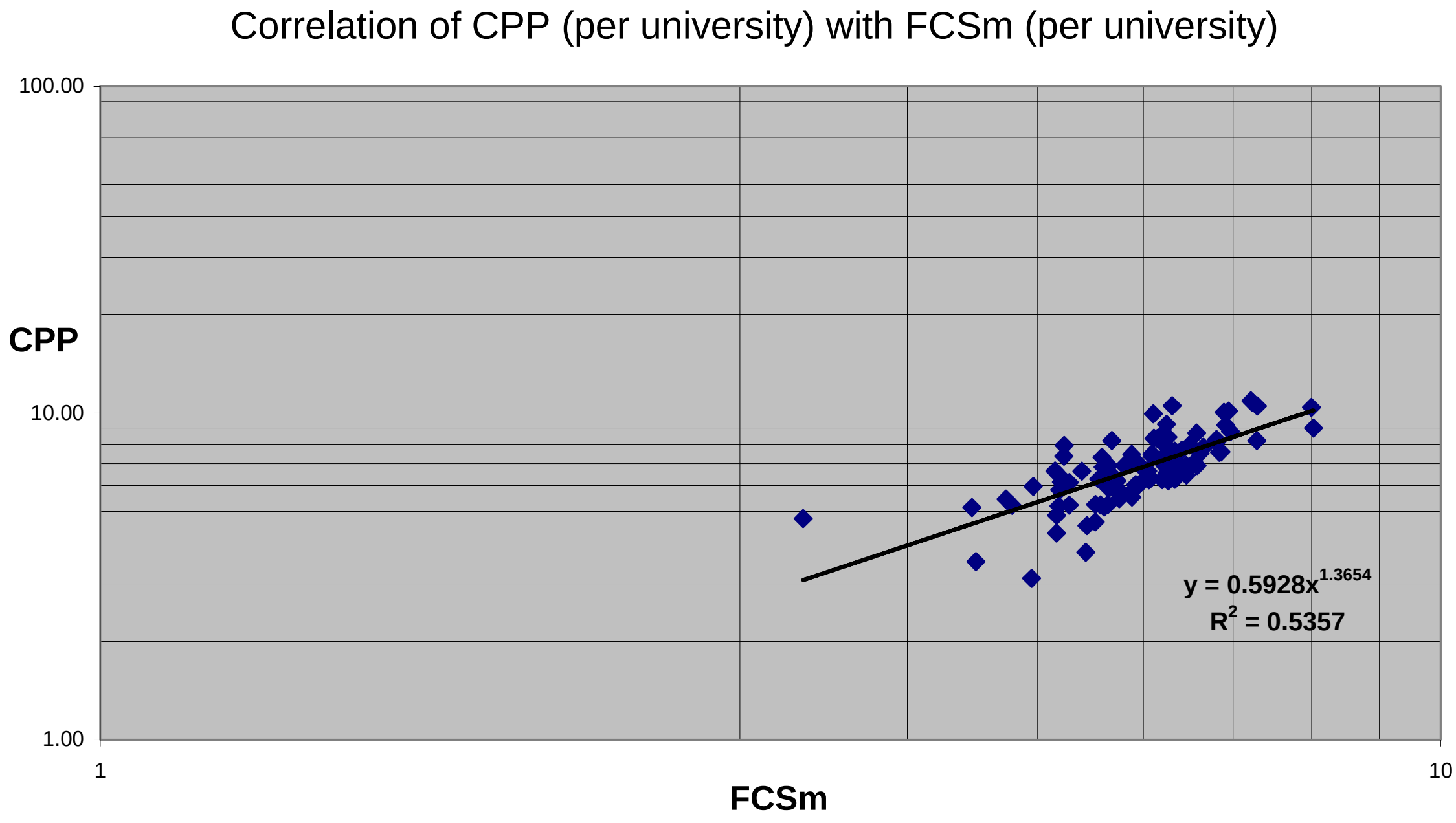

***Fig. 3.2.4:*** *Correlation of **CPP** with **FCSm** for all 100 largest European universities.*

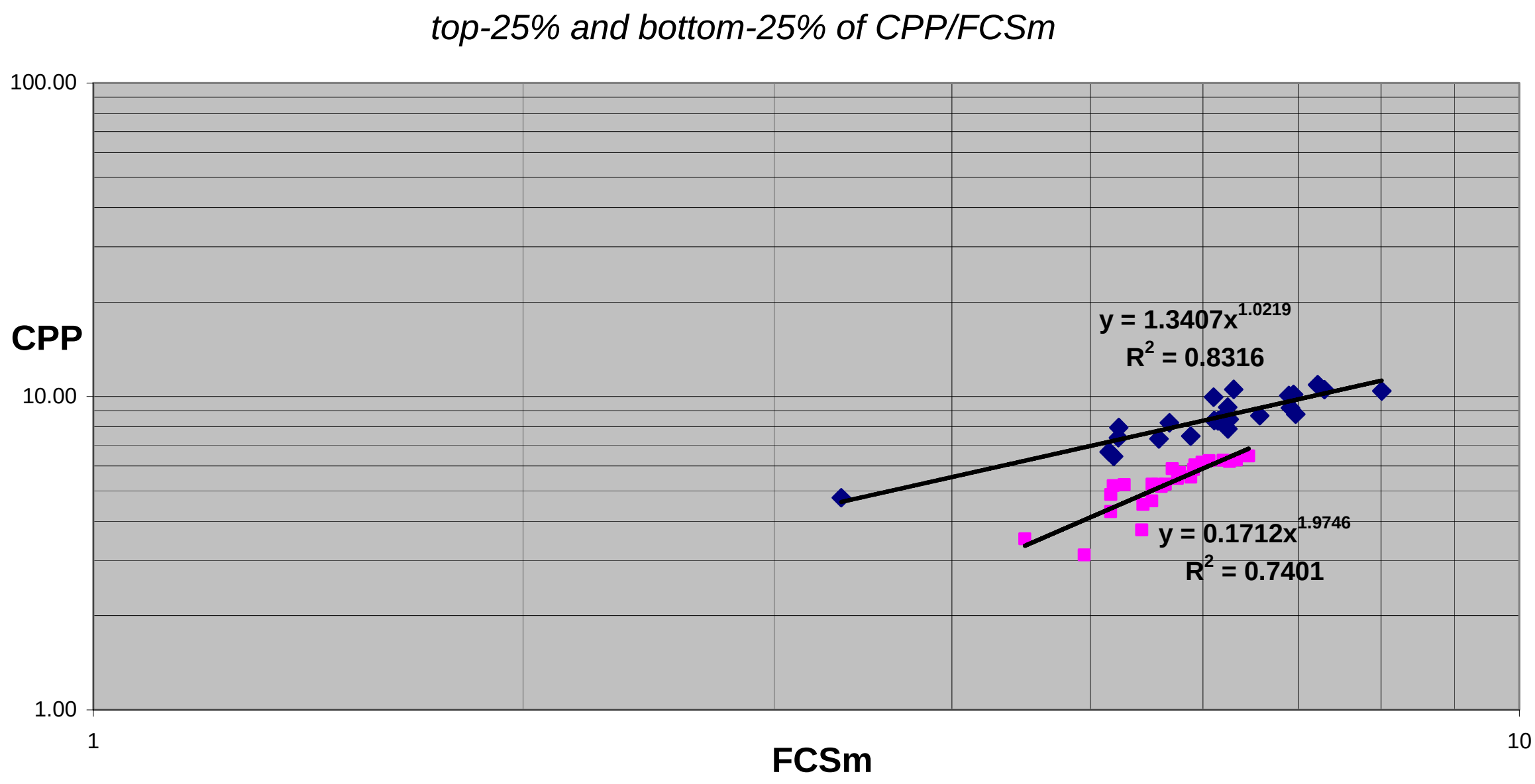

***Fig. 3.2.5:*** *Correlation of **CPP** with **FCSm** for the top-25% (diamonds) and the bottom-25% (squares) of **CPP/FCSm** distribution of the 100 largest European universities.*

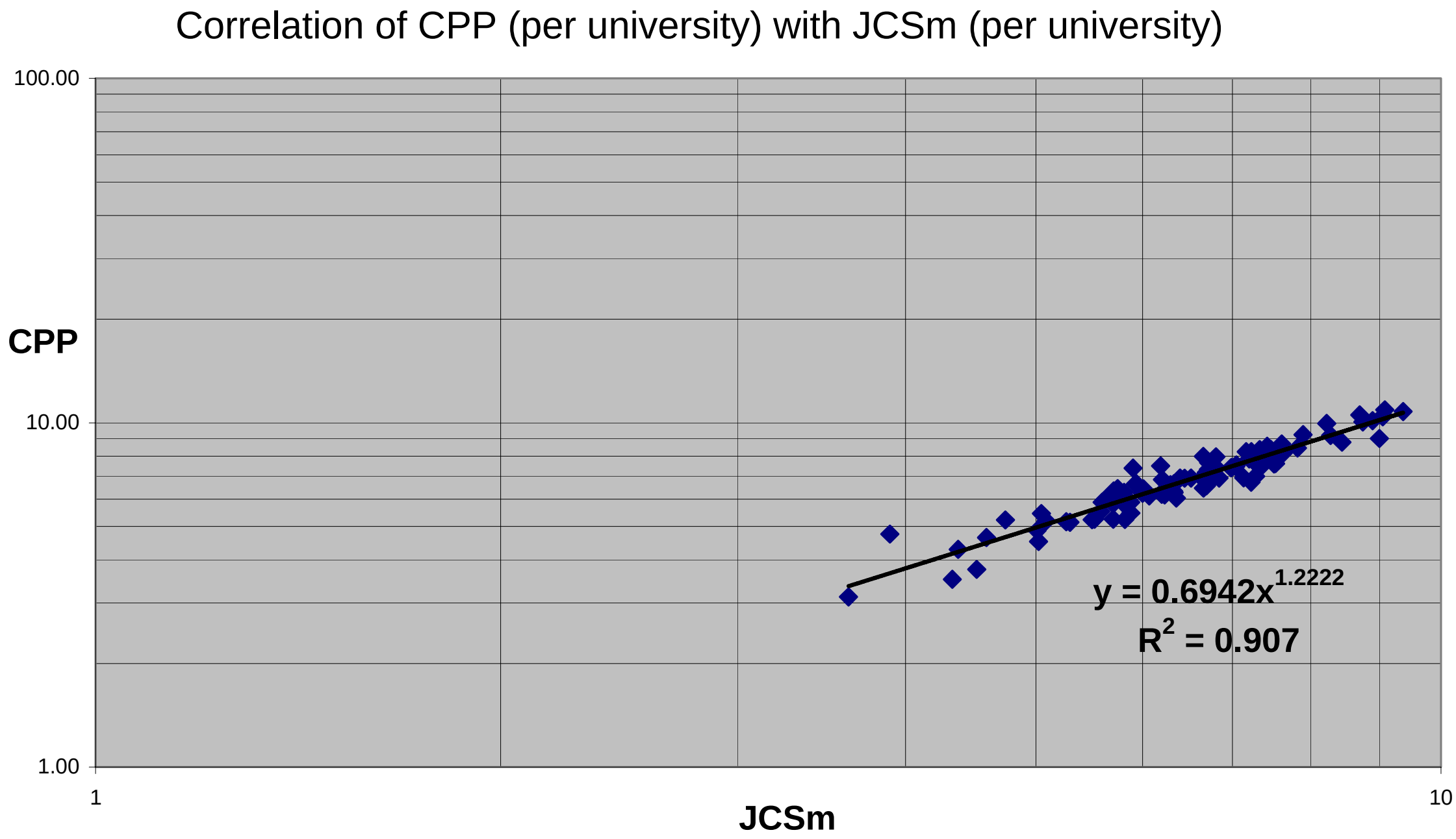

***Fig. 3.2.6**: Correlation of **CPP** with **JCSm** for all 100 largest European universities.*

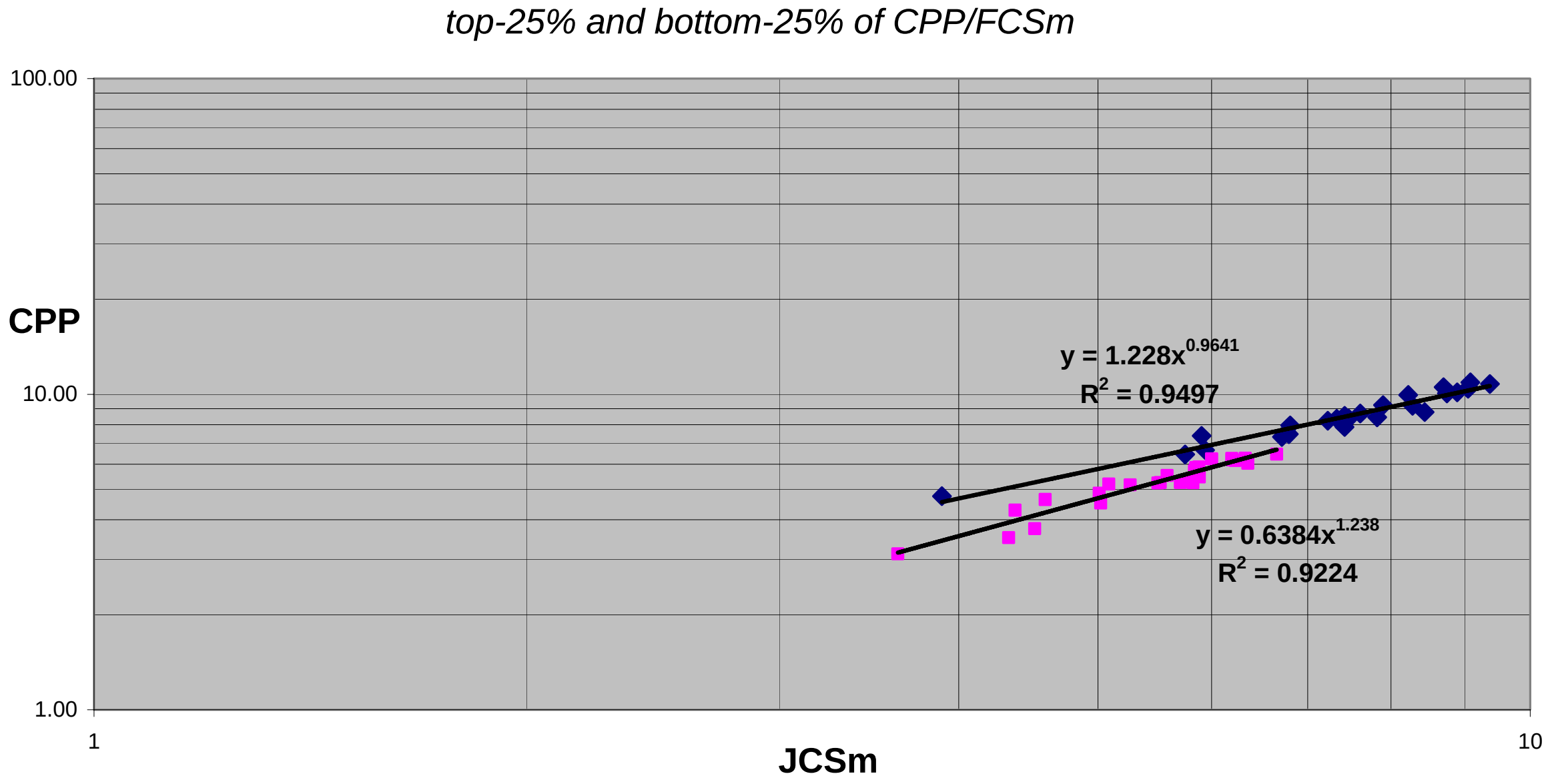

***Fig. 3.2.7:** Correlation of **CPP*** with ***JCSm** for the top-25% (diamonds) and the bottom-25% (squares) of **CPP/FCSm** distribution of the 100 largest European universities*.

Both the top- and lower-performance universities have more citations per publication (***CPP***) as a function of field citation density (***FCSm***, Fig.3.2.5) as well as of average journal impact (***JCSm***, Fig. 3.2.7). Clearly, the top universities generally have higher ***CPP*** values. We find that particularly for the lower-performance universities the field citation density (***FCSm***) provides a strong cumulative advantage in citations per publication (***CPP***) with exponent $\alpha$ = 1.97. The correlation of ***CPP*** with the average journal impact (***JCSm***) shows a less strong cumulative advantage for the lower-performance universities, ***α*** = 1.24. We also observe clearly (Fig. 3.2.7) that most top-performance universities publish in journals with significantly higher journal impact as compared to the lower performance universities. Moreover, the top-25% universities perform in terms of citations per publications (***CPP***) with a factor of about 1.3 better than the bottom-25% universities in journals with the *same* average impact. An overview of the exponents of the correlation functions is given in Table 3.2.

***Table 3.2***: *Power law exponent **α** of the correlation of **CPP** with **FCSm** and with **JCSm** for the 100 largest European universities. The differences in **α** between top- and bottom-modalities are given by **Δα**(b,t).*

| | ***FCSm*** | ***JCSm*** |
|---|---|---|
| **all** | 1.37 | 1.22 |
| | | |
| **top 25%** | 1.02 | 0.96 |
| **bottom 25%** | 1.97 | 1.24 |
| ***Δα***(*b,t*) | *0.95* | *0.28* |

Next to the impact measure ***CPP*** we also investigated the correlation of the field-normalized research performance indicator (***CPP/FCSm***) of the 100 largest European universities with field citation density and with journal impact. The results are shown for the correlation of ***CPP/FCSm*** with ***FCSm*** for the entire set of all 100 largest European universities in Fig. 3.2.8, and for the top-performance (top-25% of ***CPP/FCSm***) and lower performance universities in Fig. 3.2.9. The correlation of ***CPP/FCSm*** with ***JCSm*** for the entire set of all 100 largest European universities is presented in Fig. 3.2.10 and for the top-performance and lower performance universities in Fig. 3.2.11.

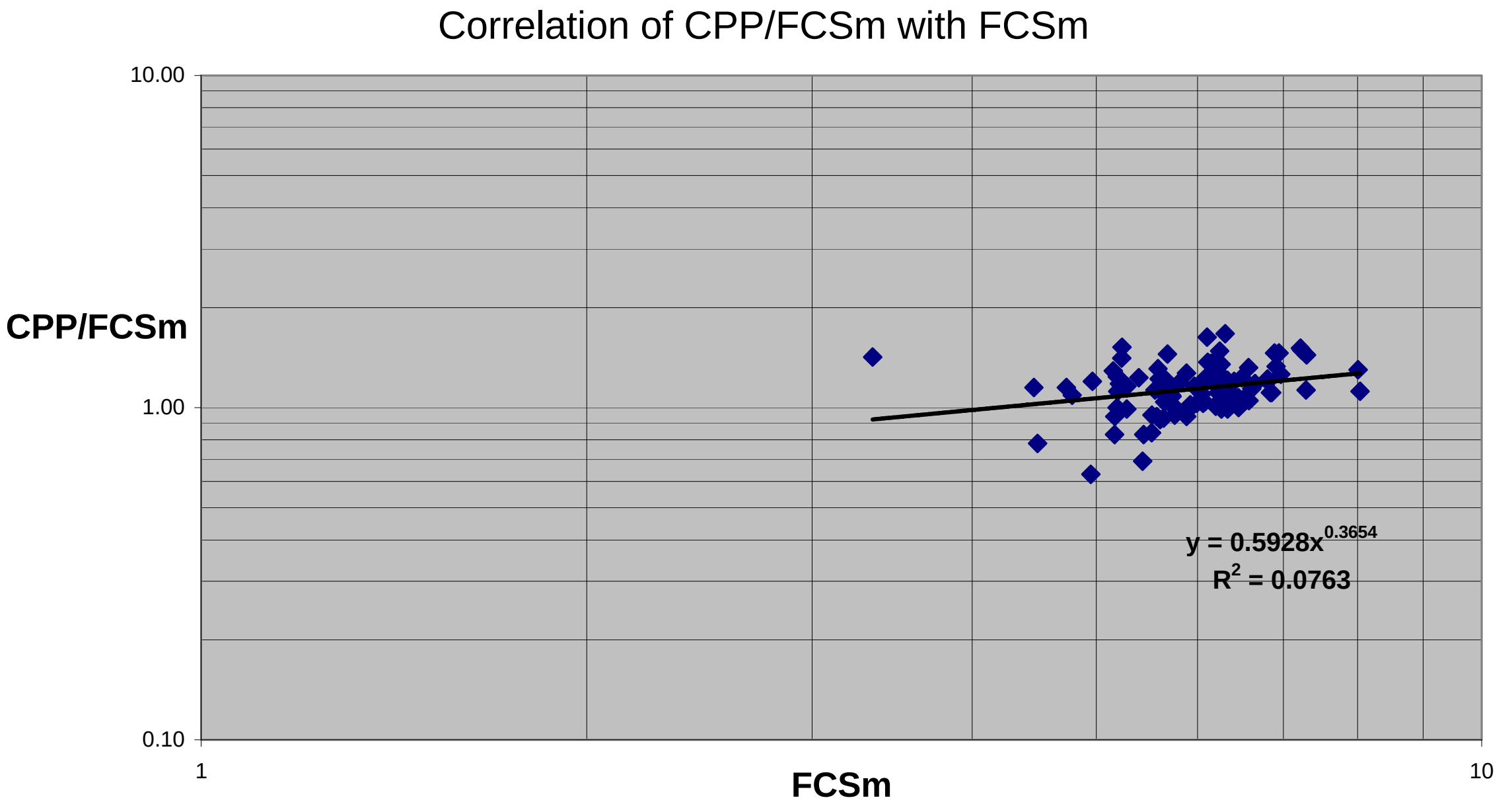

***Fig. 3.2.8:*** *Correlation of* ***CPP/FCSm*** with ***FCSm*** *for the entire set of the 100 largest European universities.*

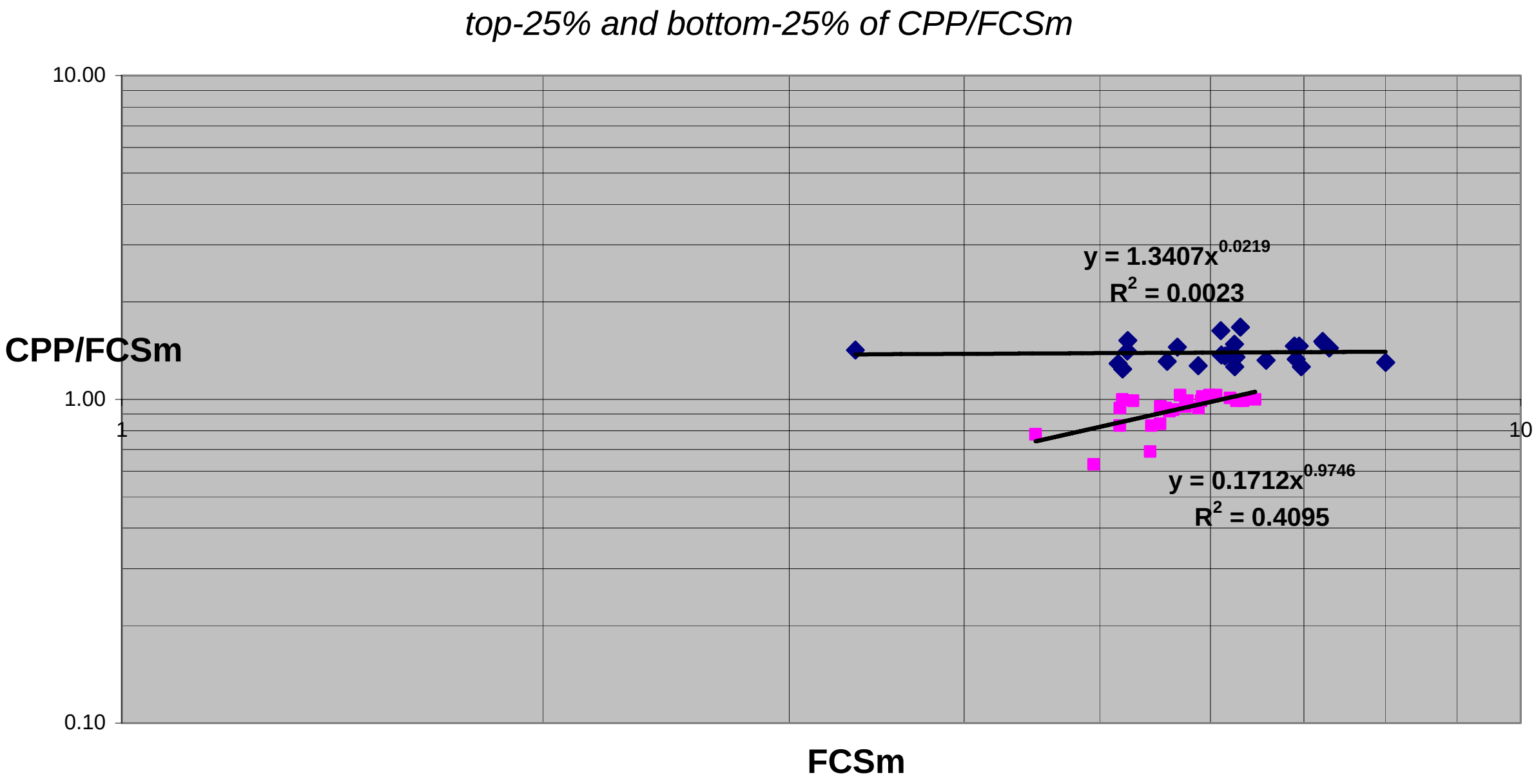

***Fig. 3.2.9:*** *Correlation of* ***CPP/FCSm*** with ***FCSm*** *for the top-25% (diamonds) and the bottom-25% (squares) of* ***CPP/FCSm*** *distribution of the 100 largest European universities.*

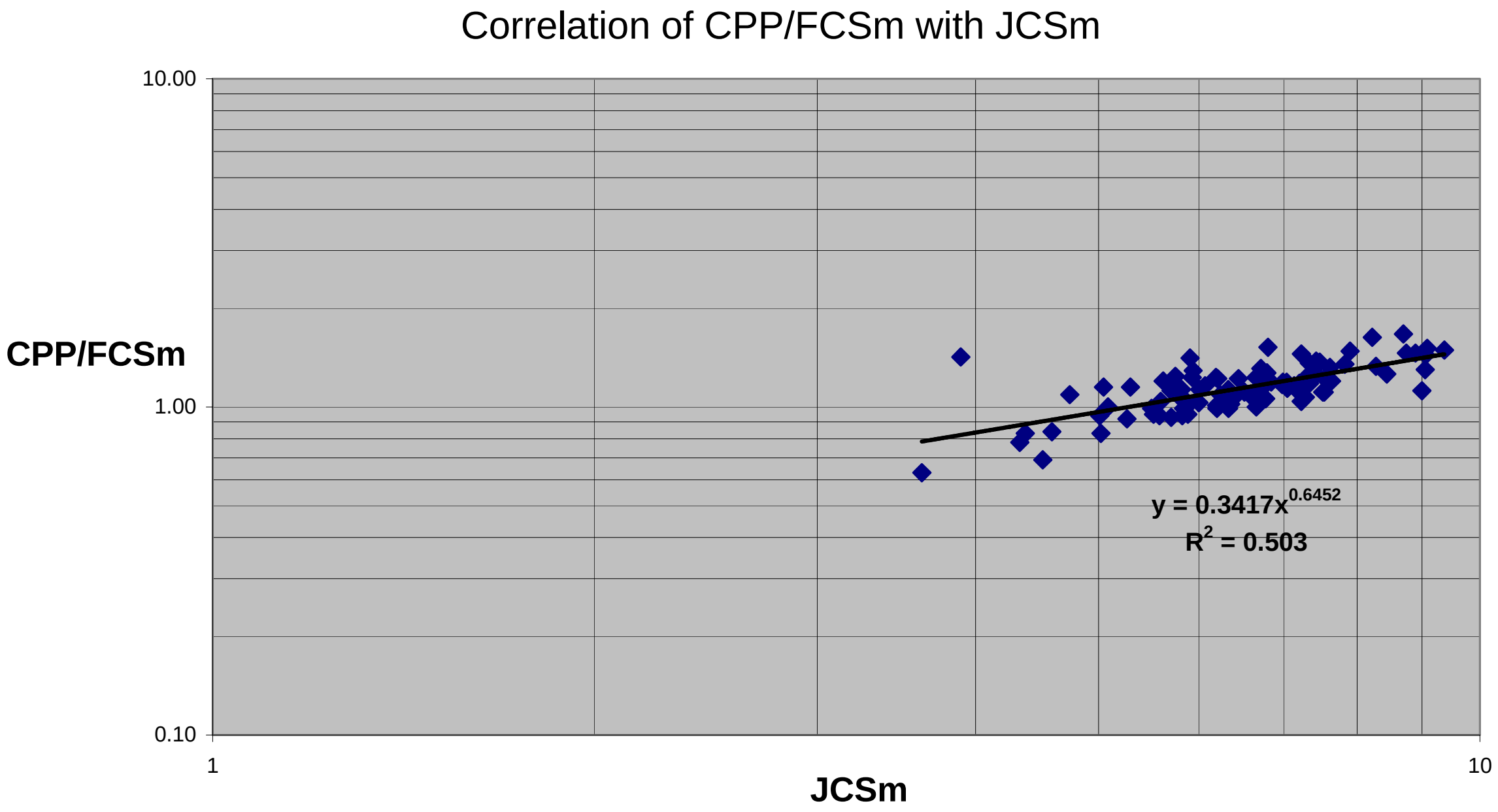

***Fig. 3.2.10:*** *Correlation of* ***CPP/FCSm*** with ***JCSm*** *for the entire set of the 100 largest European universities.*

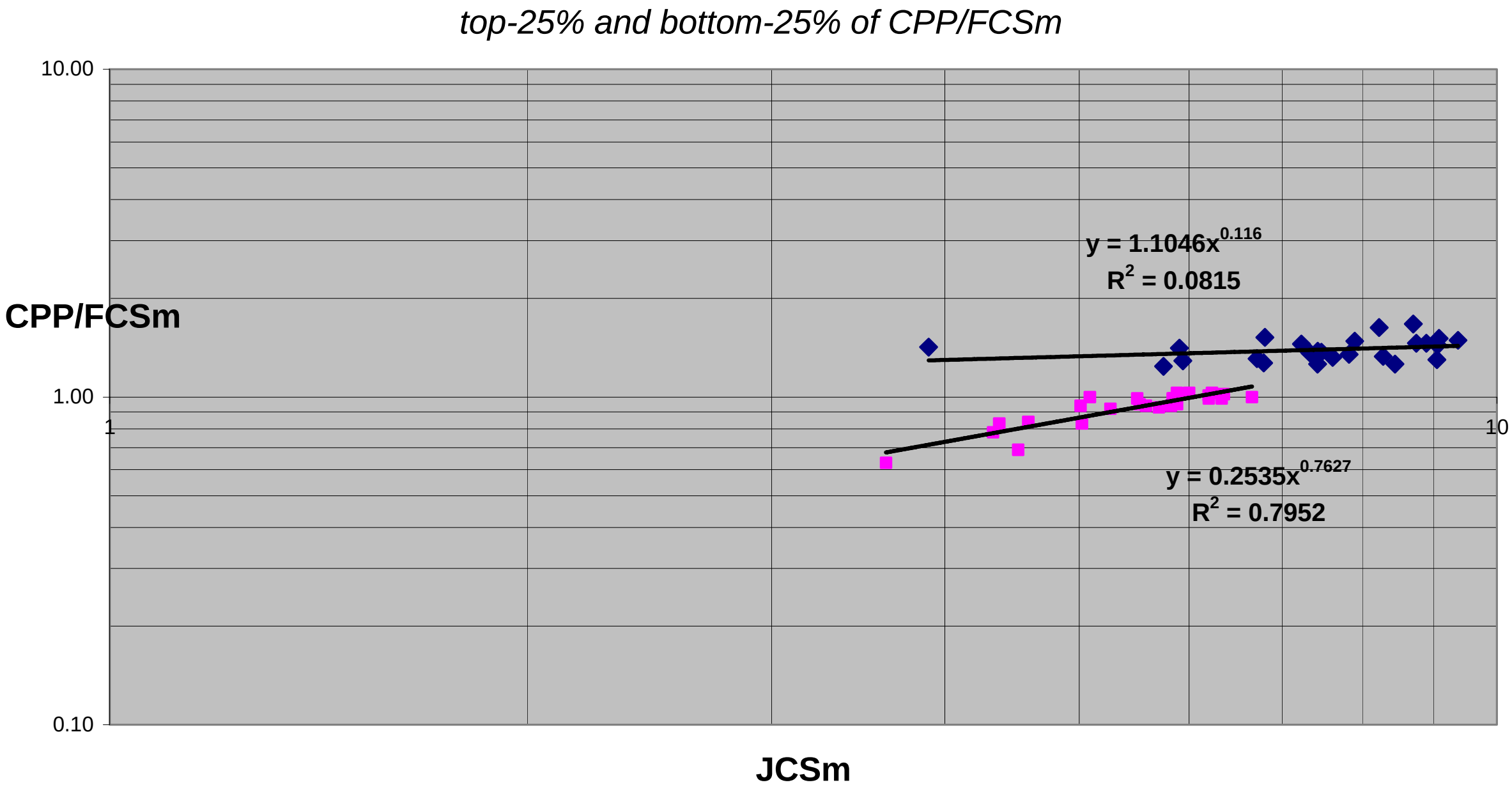

***Fig. 3.2.11:*** *Correlation of* ***CPP/FCSm*** with ***JCSm*** *for the top-25% (diamonds) and the bottom-25% (squares) of* ***CPP/FCSm*** *distribution of the 100 largest European universities.*

We observe that the research performance of the top universities is independent of field citation density (***FCSm***). For the lower-performance universities there is a slight increase of performance as a function of ***FCSm***. The results for the average journal impact (***JCSm***) are similar but more outspoken. Again we notice that top-performance universities have a strong preference for the higher-impact journals.

Finally, we analysed the correlation between the number of not-cited publications (***Pnc***) of a university and its average journal impact level (***JCSm***). The results are shown in Fig. 3.2.12 for the entire set of 100 universities and in Fig. 3.2.13 for the top- en lower-performance universities. We see a quite significant correlation between these two variables. Very clearly the top universities have the lowest ***Pnc***. Given the strong correlation between ***CPP*** and ***JCSm*** (see Fig. 3.2.6) we can also expect a significant correlation between ***Pnc*** and ***CPP***, as confirmed nicely by Fig. 3.2.14 for the entire set of 100 universities and in Fig. 3.2.15 for the top- en lower-performance universities. Thus, we find that the higher the average journal impact of the publications of a university, the lower the number of not-cited publications. Also, the higher the average number of citation per publication in a university, the lower the number of not-cited publications. In other words, universities that are cited more per paper also have more cited papers. These findings underline the generally good correlation at the university level between the average number of citations per publication in a university, and its average journal impact.

We also find that the relation between the relative number of not-cited publications (***Pnc***) and the mean number of citations per publication *(**CPP**)* can be written in good approximation as

$$Pnc = 1/\sqrt{(CPP)}.$$

This expression reflects the characteristics of the citation-distribution function as it is the relation between the number of publications with zero citations and the average number of citations per publications.

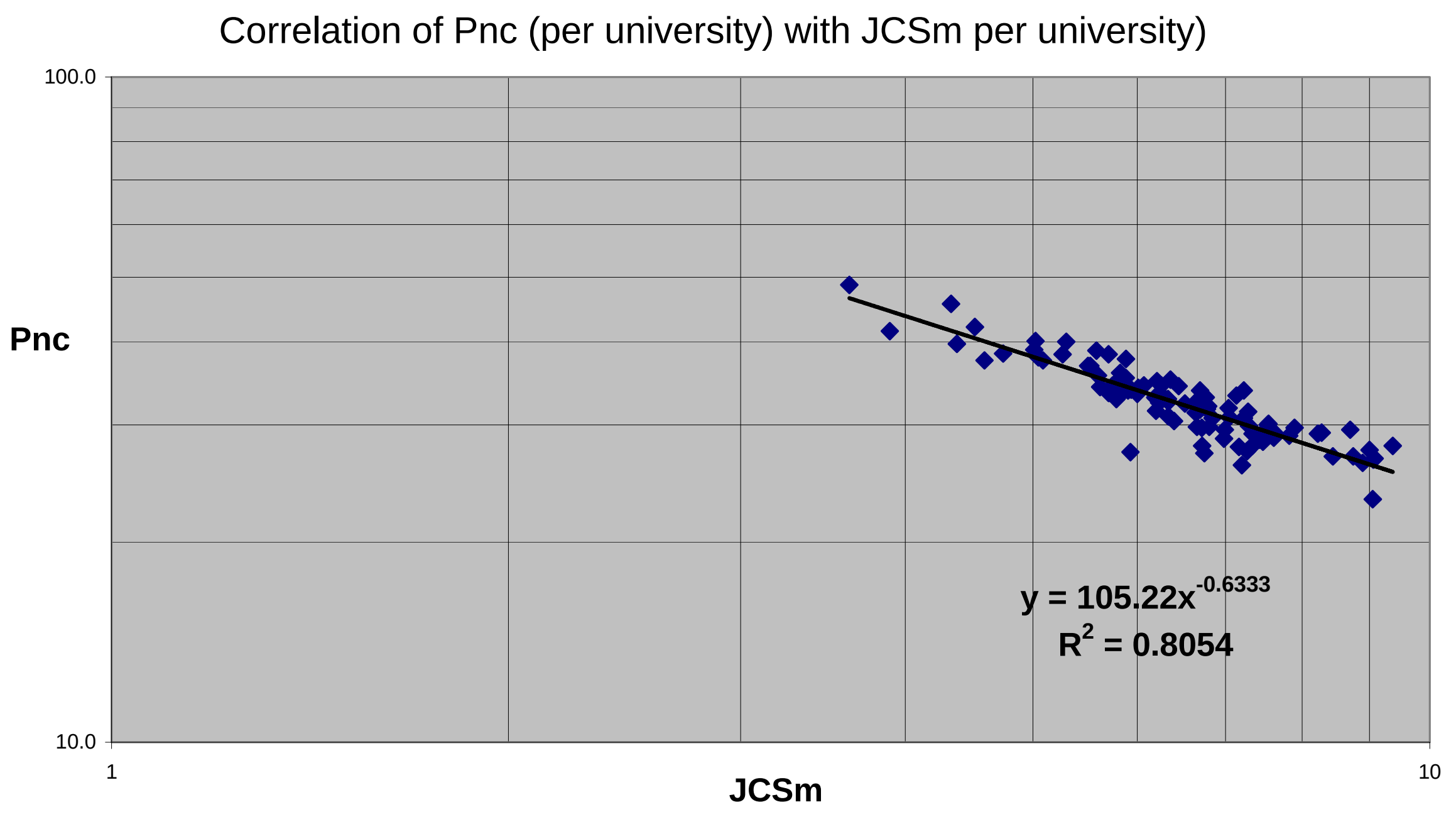

***Fig. 3.2.12***: *Correlation of the relative number of not cited publications (**Pnc**) with the mean journal impact (**JCSm**) of the 100 largest European universities.*

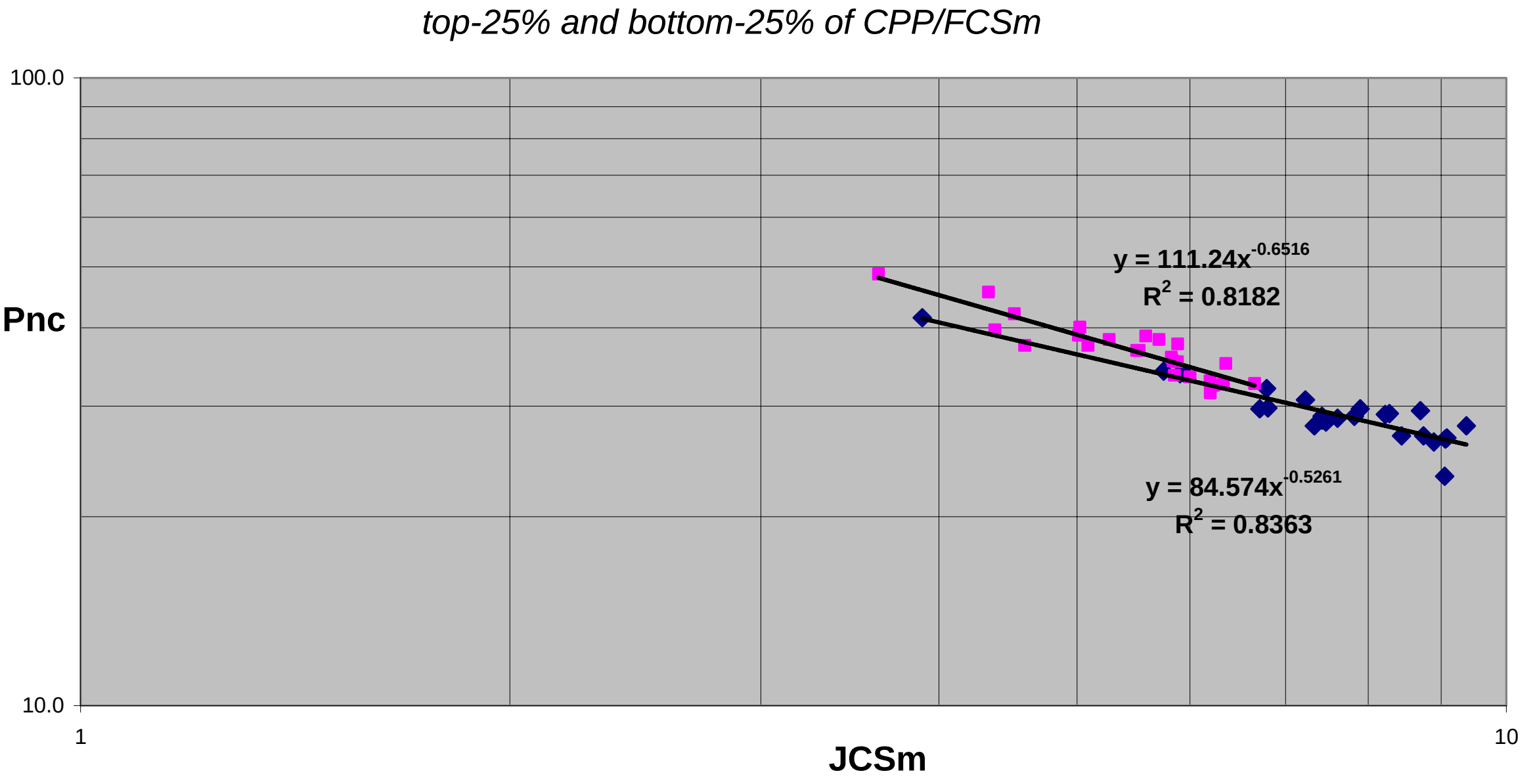

***Fig. 3.2.13:*** *Correlation of the relative number of not cited publications (**Pnc**) with the mean journal impact (**JCSm**) for the top-25% (of **CPP/FCSm**) universities (diamonds), and the bottom-25% universities (squares).*

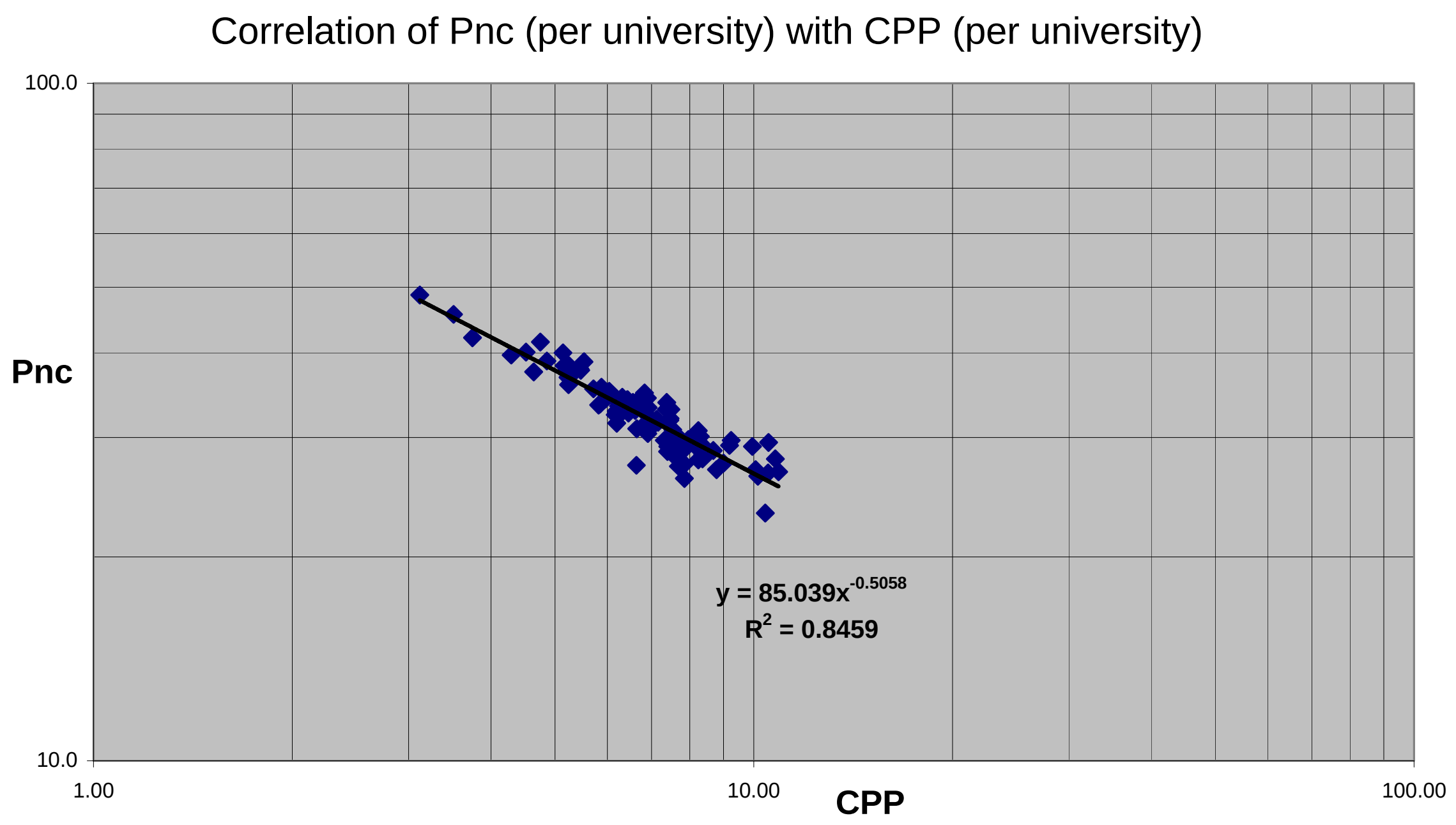

***Fig. 3.2.14****: Correlation of the relative number of not cited publications (****Pnc****) with the mean number of citations per publication (****CPP****) of the 100 largest European universities*.

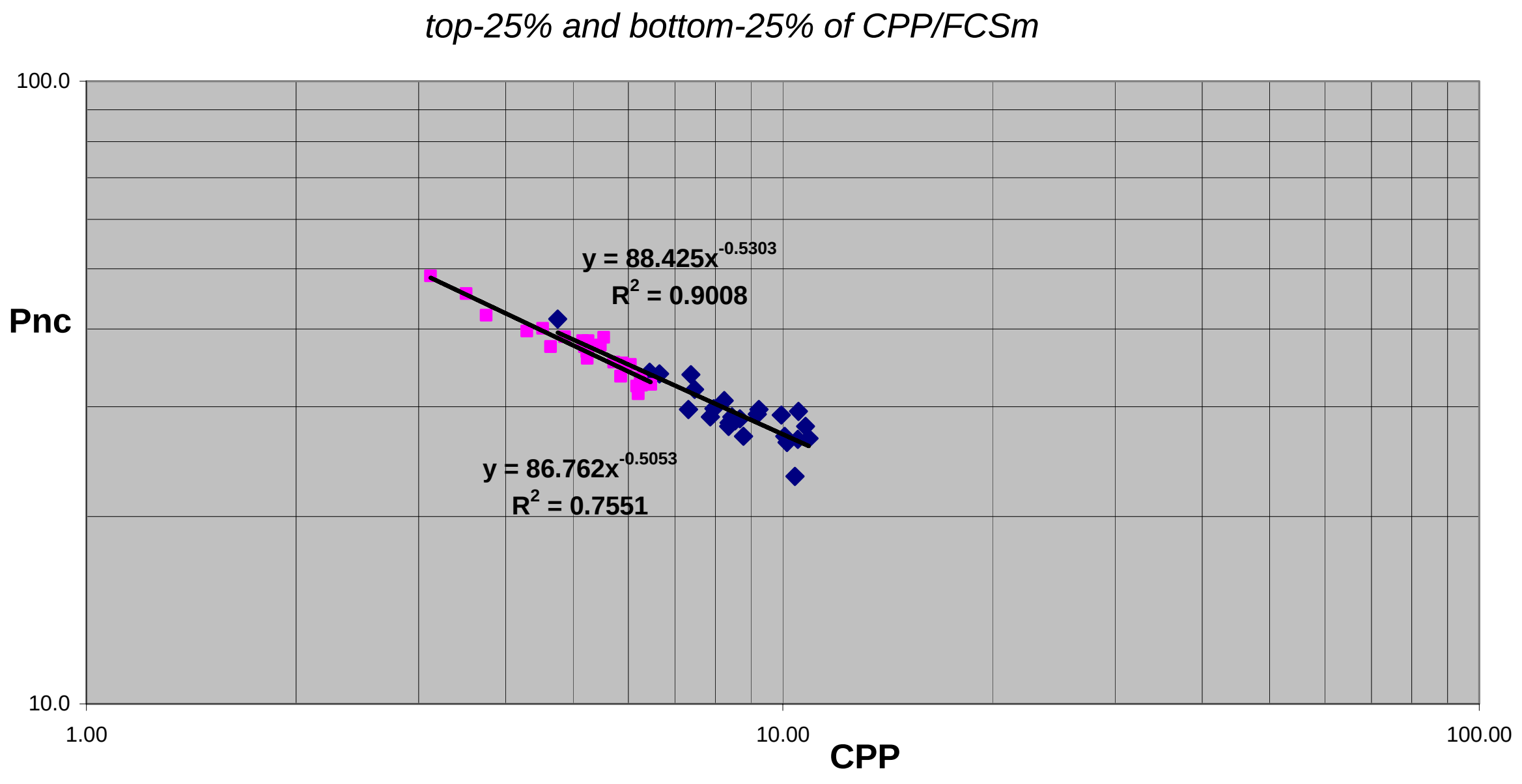

***Fig. 3.2.15:*** *Correlation of the relative number of not cited publications (****Pnc****) with the mean number of citations per publication (****CPP****) for the top-25% (of* ***CPP/FCSm****) universities (diamonds), and the bottom-25% universities (squares)*.

### 3.3 Characteristics of self-citations

In this section we present a first analysis of a specific feature of the science system, the statistical properties of self-citations. We calculated the correlation between size (the total number of publications ***P***) and the total number of citations ***C*** for all 100 largest European universities. Fig. 3.3.1 shows that this correlation is described with high significance by a power law:

$$Sc(P) = 0.53\, P^{1.15} .$$

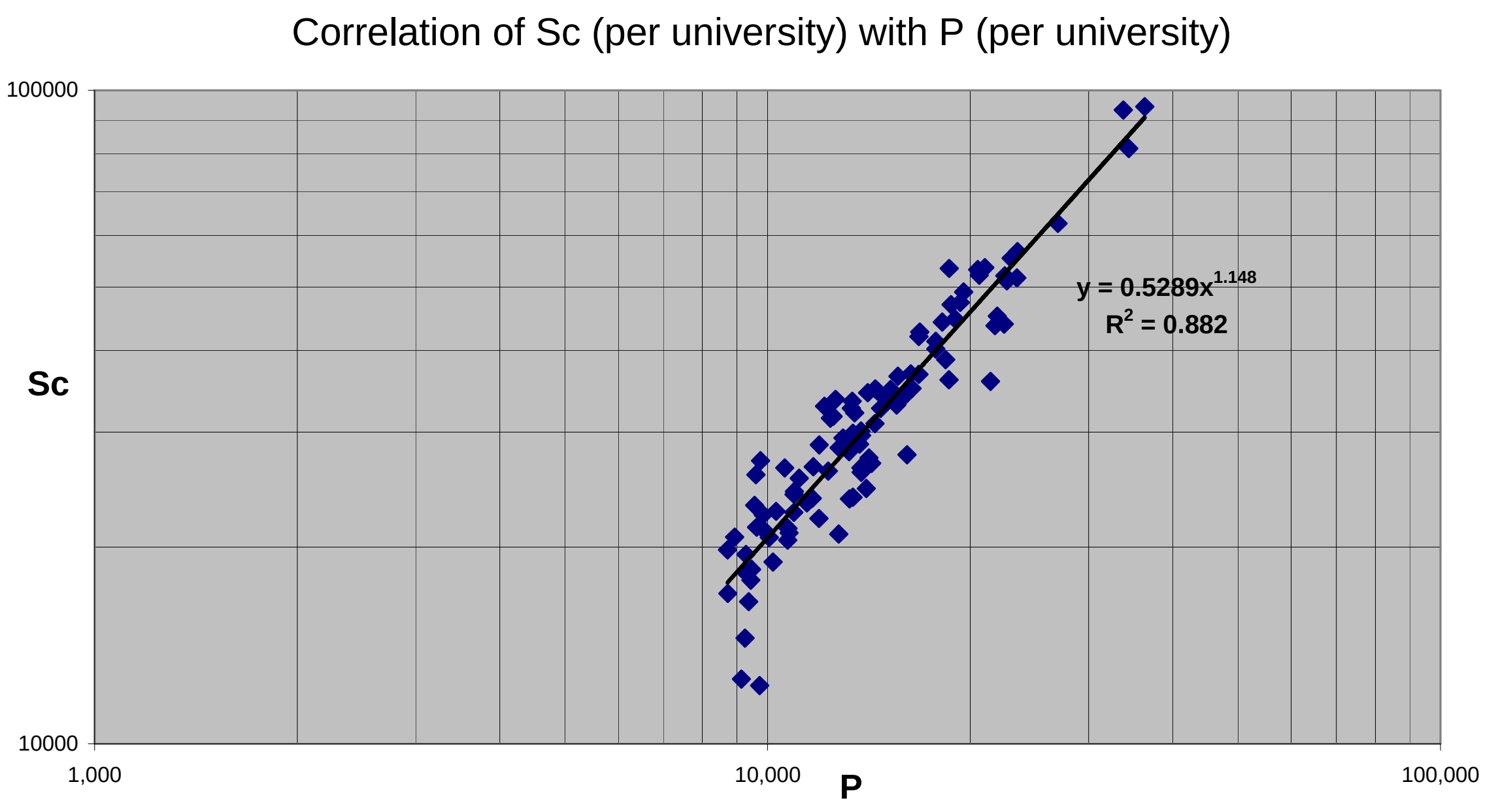

***Fig. 3.3.1:*** *Correlation of the number of self-citations (**Sc**) received per university with the number of publications (**P**) of these universities, for all 100 largest European universities.*

At the lower side of ***P*** (and ***Sc***) we again observe the 'outliers' as in the case of the (external) citations (Fig. 3.1.1). We find that the size of universities leads to a cumulative advantage (with exponent $\alpha$=+1.15) for the number of self-citations given by these universities. Gradual differentiation between top- and lower-performance (top/bottom 10%, 25%, and 50%) enables us to study the correlation of ***Sc*** with ***P*** in more detail as presented in Figs. 3.3.2 - 3.3.4. We see that the group of highest performance universities (top-10%) does not have a cumulative advantage (exponent around 1), whereas the bottom-10% exponent is heavily determined by the outliers. The broader top-25% and the bottom-25% show a slight cumulative advantage ($\alpha$= 1.11 and 1.15, respectively). If we divide the entire set of universities in a top- and bottom-50% we see that both subsets have more or less equal exponents (around 1.11).

In Fig. 3.3.5 we show that the fraction (percentage) of self-citations (***%Sc***) decreases slightly with size (***P***), but this correlation is not very significant. More significant is the decrease of the fraction of self-citations as a function of research performance ***CPP/FCSm***, as shown in Fig. 3.3.6. We also observe a clear decrease of self-citations for the 100 largest universities in Europe as a function of average field citation density ***FCSm***, Fig. 3.3.7, average journal impact ***JCSm***, Fig. 3.3.8, and of field-normalized journal impact ***JCSm/FCSm***, see Fig. 3.3.9.

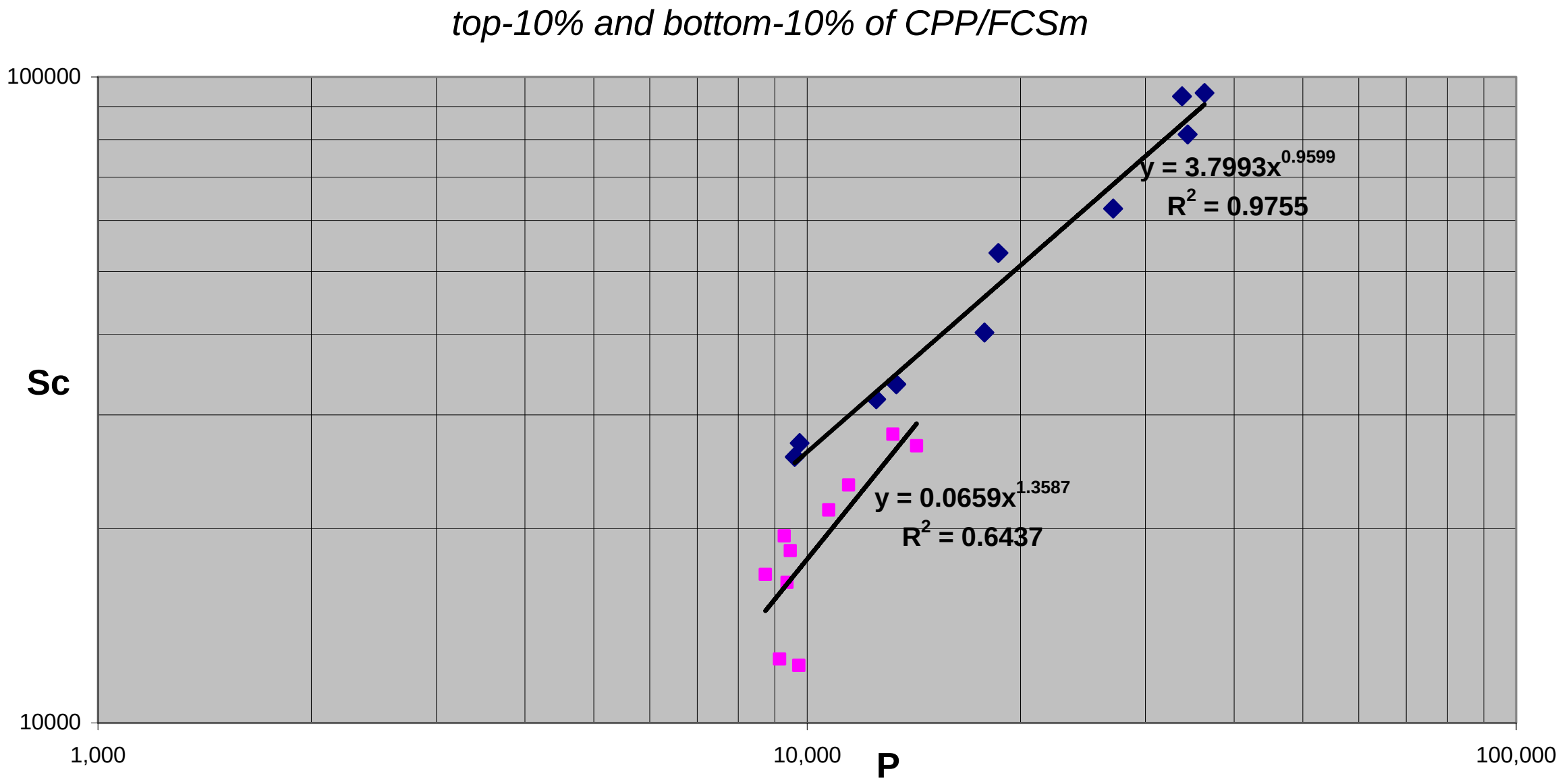

***Fig. 3.3.2:*** *Correlation of the number of self-citations (**Sc**) received per university with the number of publications (**P**), for the top-10% (of **CPP/FCSm**) universities (diamonds), and the bottom-10% universities (squares) with the 100 largest European universities.*

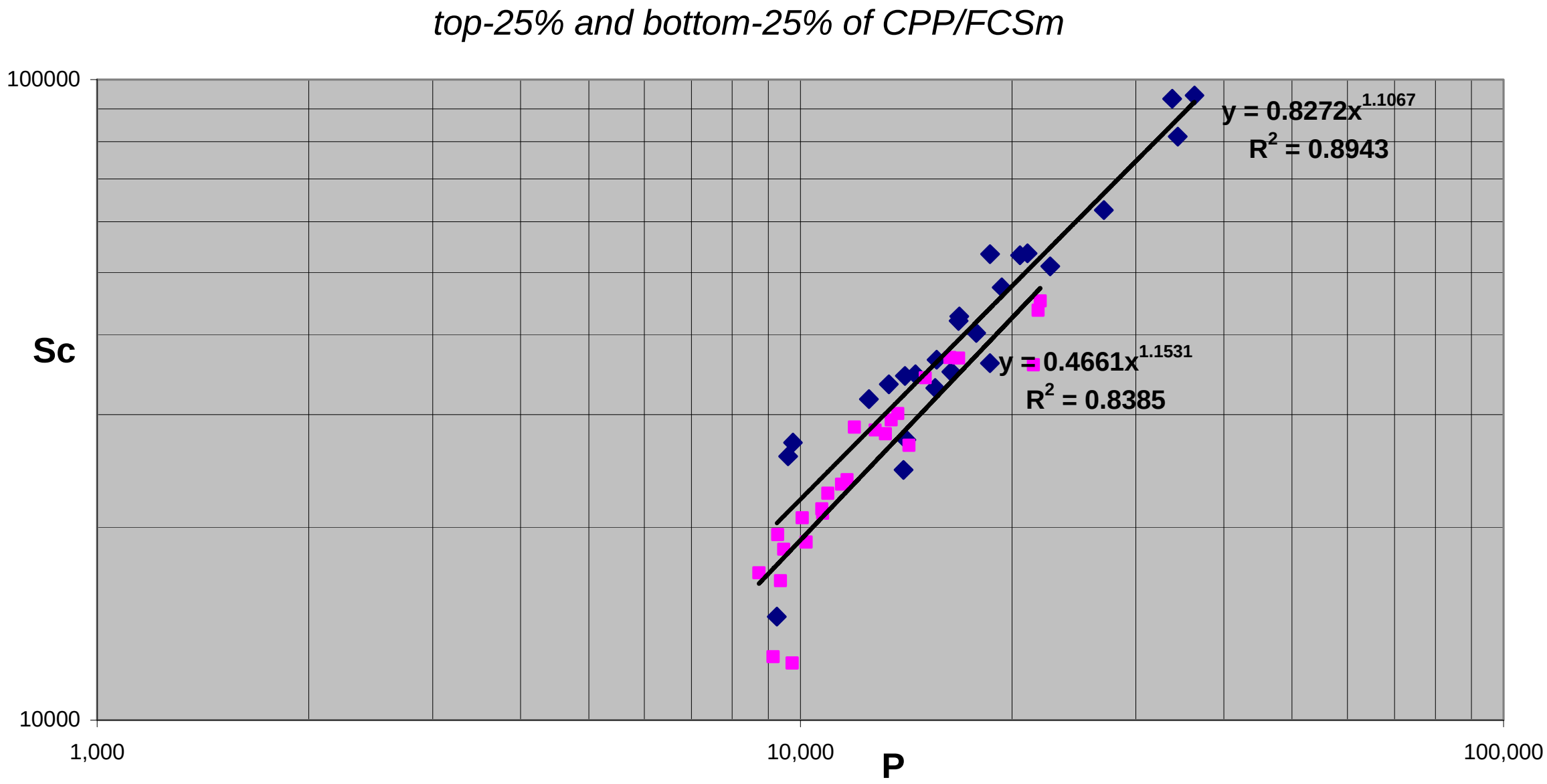

***Fig. 3.3.3:*** *Correlation of the number of self-citations (**Sc**) received per university with the number of publications (**P**), for the top-25% (of **CPP/FCSm**) universities (diamonds), and the bottom-25% universities (squares) with the 100 largest European universities.*

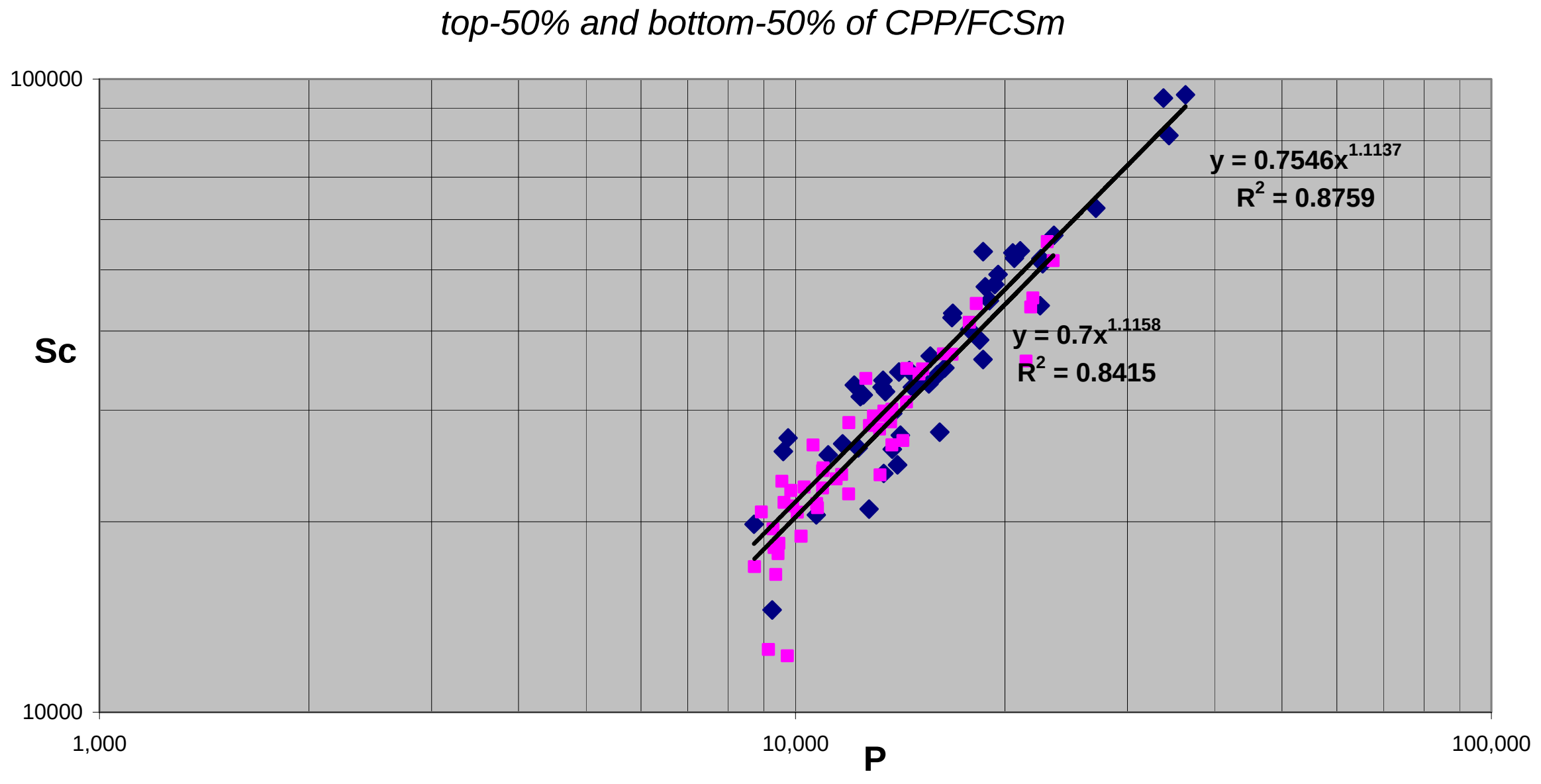

***Fig. 3.3.4:*** *Correlation of the number of self-citations (**Sc**) received per university with the number of publications (**P**), for the top-50% (of **CPP/FCSm**) universities (diamonds), and the bottom-50% universities (squares) with the 100 largest European universities.*

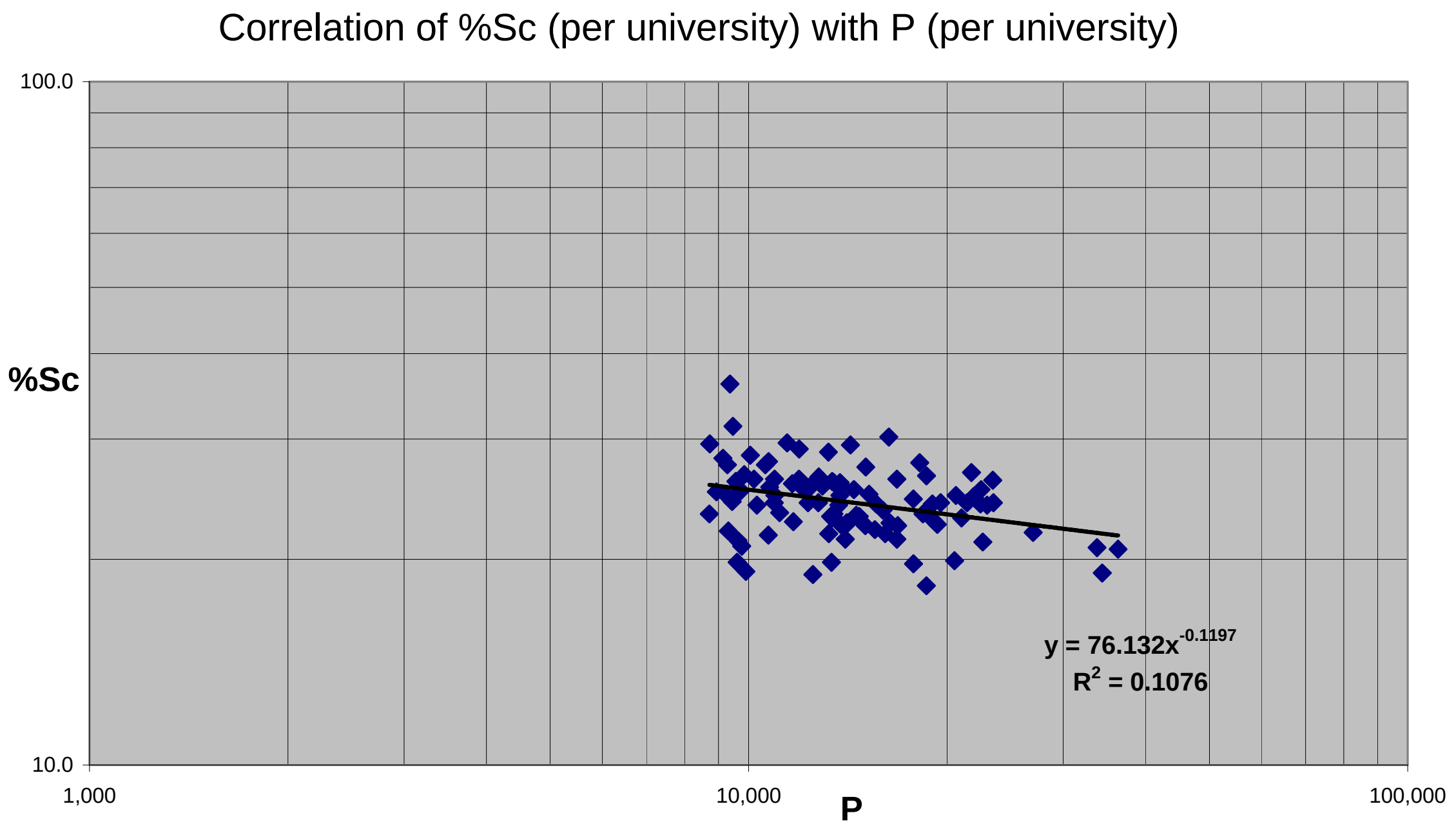

***Fig. 3.3.5:*** *Correlation of the relative number of self-citations (%**Sc**) per university with the number of publications (**P**) of these universities, for all 100 largest European universities*.

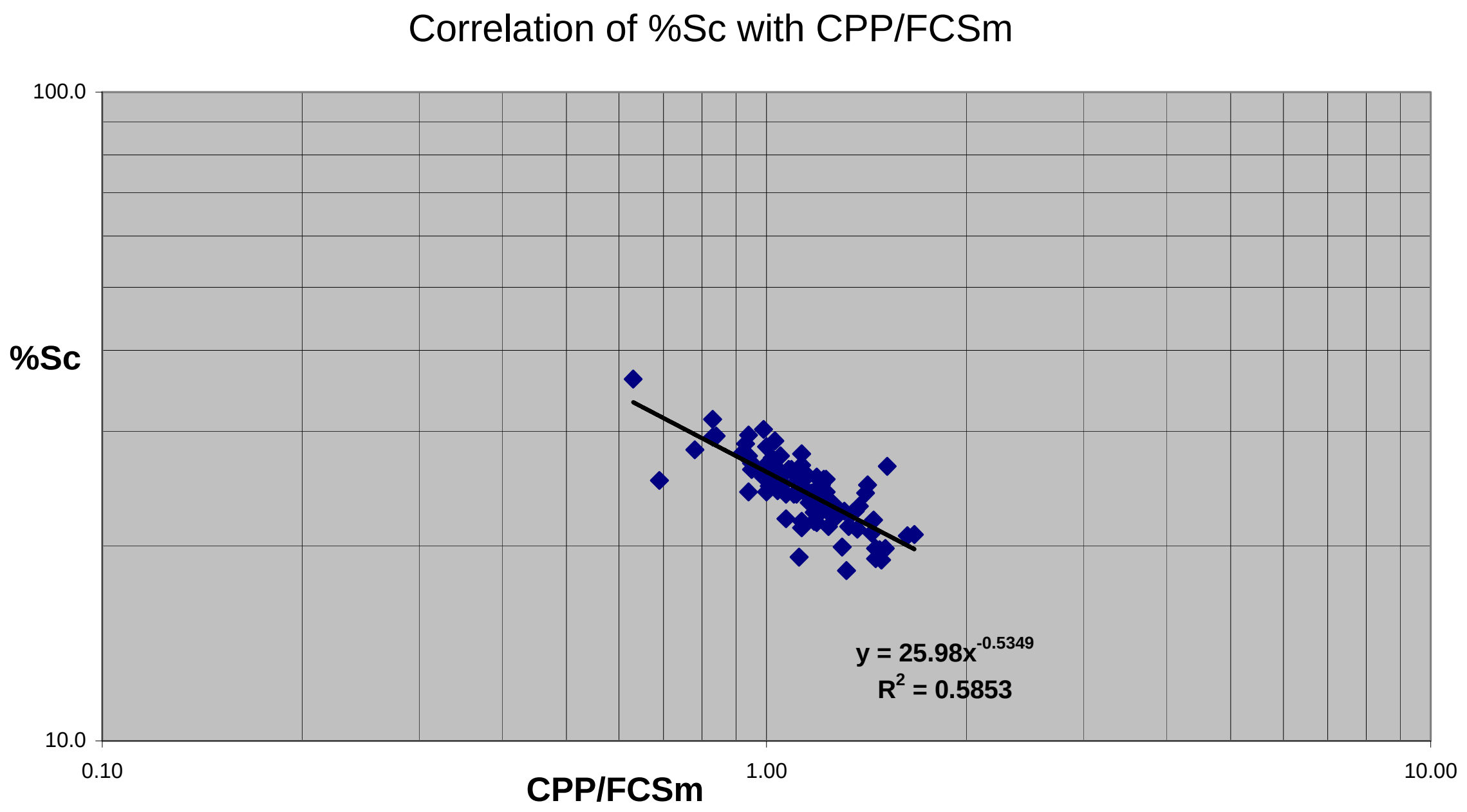

***Fig. 3.3.6:*** *Correlation of the relative number of self-citations (%**Sc**) per university with the performance (**CPP/FCSm**) of these universities, for all 100 largest European universities.*

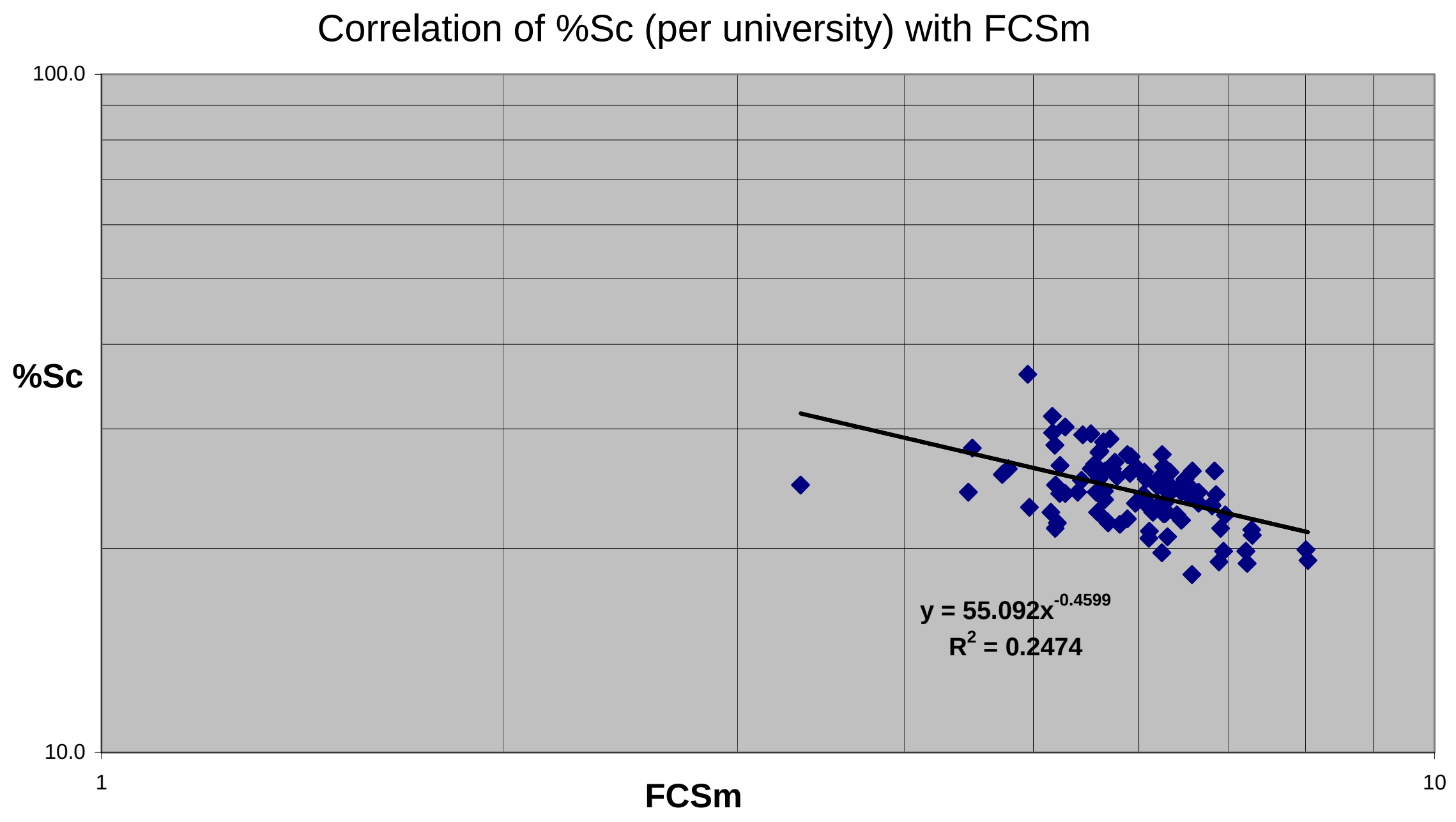

***Fig. 3.3.7:*** *Correlation of the relative number of self-citations (%**Sc**) per university with the field citation density (**FCSm**) of these universities, for all 100 largest European universities.*

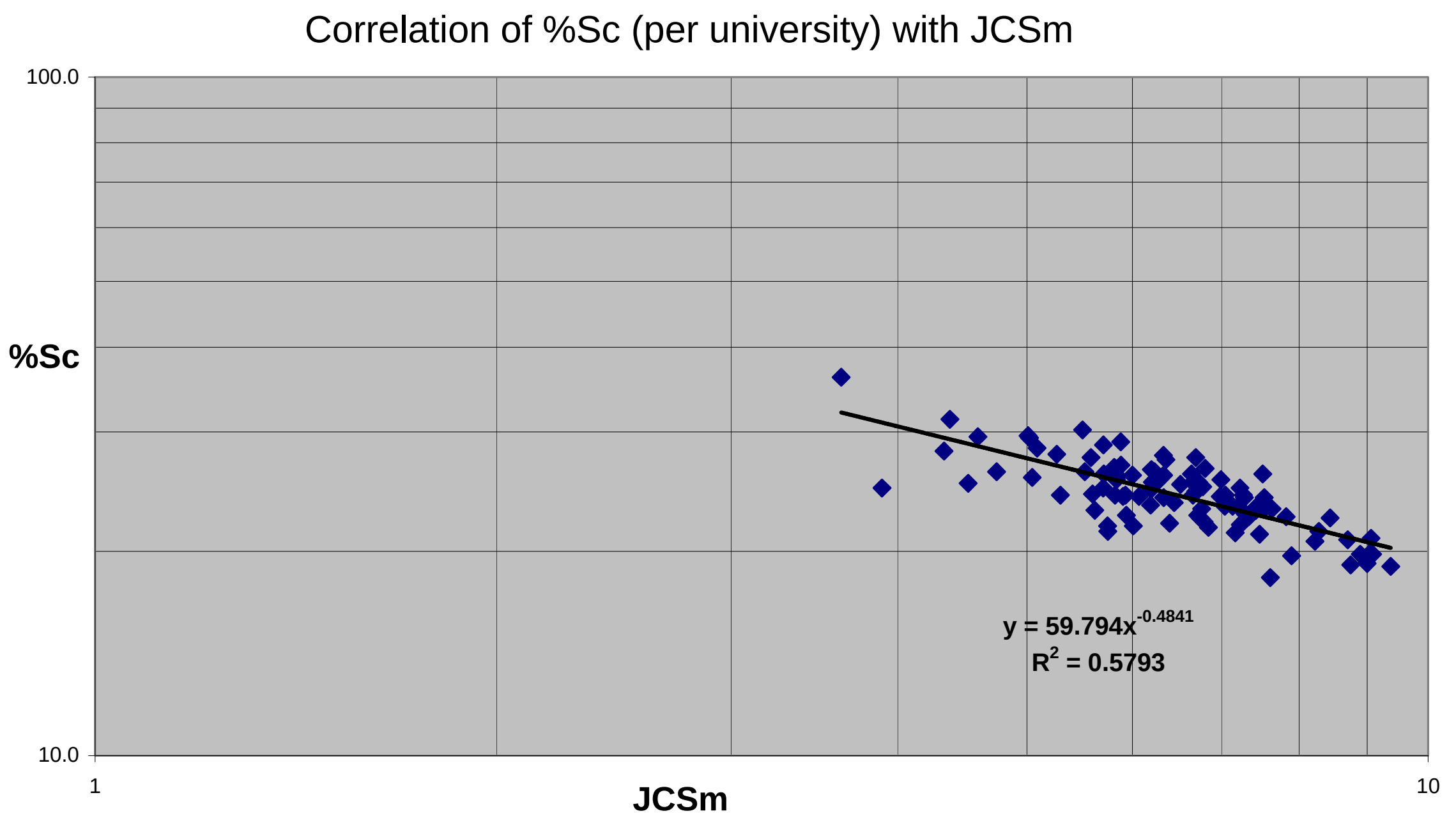

***Fig. 3.3.8:*** *Correlation of the relative number of self-citations (%**Sc**) per university with the average journal impact (**JCSm**) of these universities, for all 100 largest European universities.*

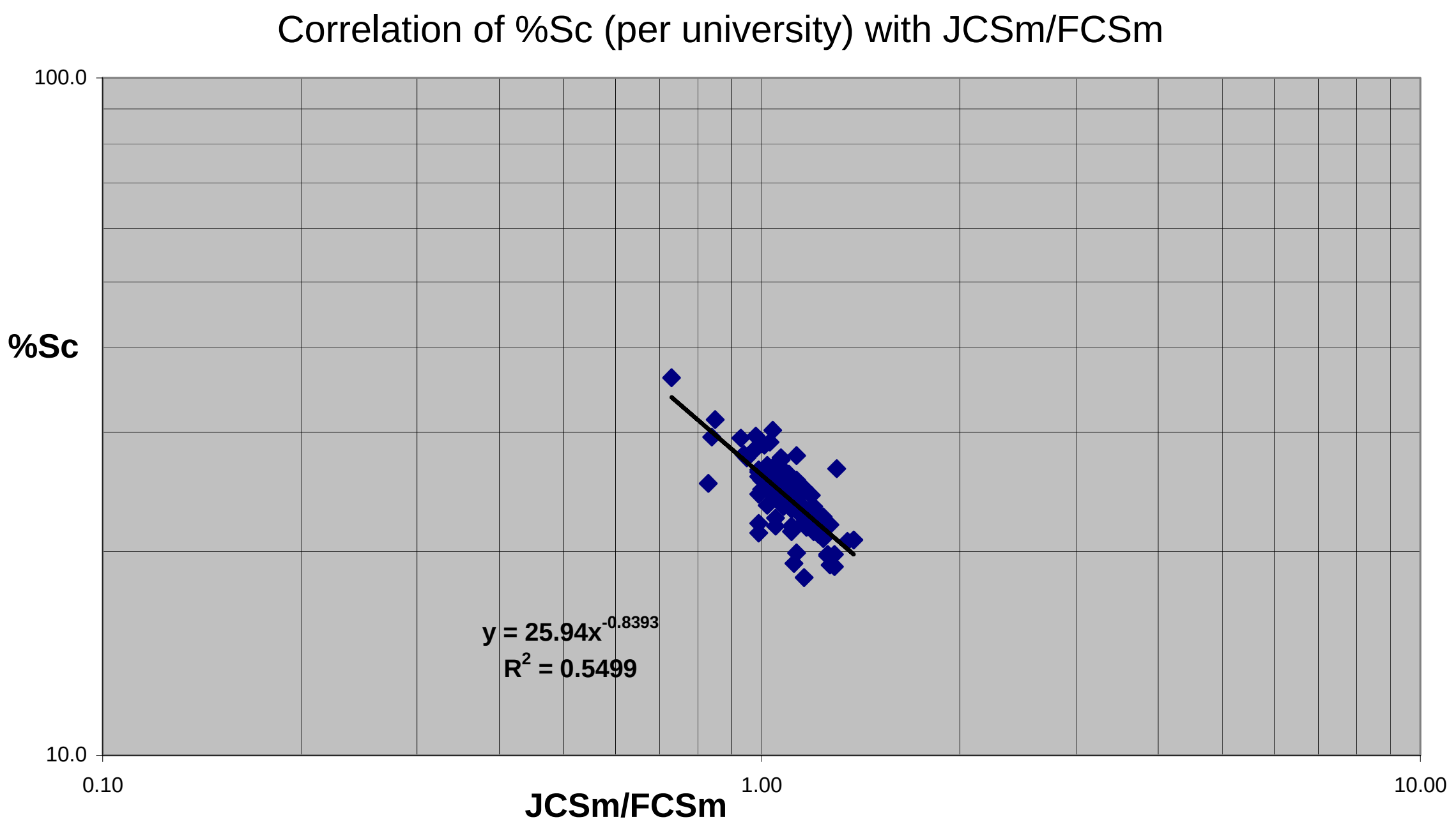

***Fig. 3.3.9:*** *Correlation of the relative number of self-citations (%**Sc**) per university with the field-normalized journal impact (**JCSm/FCSm**) of these universities, for all 100 largest European universities.*

## 4. Summary of the main findings and concluding remarks

For the 100 largest European universities we studied statistical properties of bibliometric characteristics related to research performance, field citation density and journal impact. Our five main observations are as follows.

First, we find a size-dependent cumulative advantage for the impact of universities in terms of total number of citations. Quite remarkably, lower performance universities have a larger size-dependent cumulative advantage for receiving citations than top-performance universities. We found in previous work a similar scaling rule at the level of research groups and therefore we conjecture that this scaling rule is a prevalent property of the science system. We also observe that the top universities are about twice as efficient in receiving citations (***C***) as compared to the bottom-performance universities. Our criterion of top- or low performance is based on the field-normalized indicator ***CPP/FCSm***. We hypothesize that in network terms this indicator represents the 'fitness' of a university as a node in the science system. It brings a university in a better position to acquire additional links (in terms of citations) on the basis of quality (high performance).

Second, we find that for the lower-performance universities the fraction of not-cited publications decreases with size. We explain this phenomenon with a model in which size is advantageous in an 'internal promotion mechanism' to get more publications cited. Thus, in this model size is a distinctive parameter which acts as a bridge between the macro-picture (characteristics of the entire set of universities) and the micro-picture (characteristics within a university). We find that the higher the average journal impact of a university, the lower the number of not-cited publications. Also, the higher the average number of citations per publication in a university, the lower the number of not-cited publications. In other words, universities that are cited more per paper also have more cited papers.

Third, we find that the average research performance of university measured by our crown indicator ***CPP/FCSm*** does not 'dilute' with increasing size. Apparently large universities, particularly the top-performance universities are characterized by 'big and beautiful'. In other words, they succeed in keeping a high performance over a broad range of activities. This most probably is an indication of their overall scientific and intellectual attractive power.

Fourth, we observe that particularly the low field citation density and the low journal impact universities have a considerably size-dependent cumulative advantage for the total number of citations. We find that particularly for the lower-performance universities the field citation density (***FCSm***) provides a strong cumulative advantage in citations per publication (***CPP***). We also observe clearly that most top-performance universities publish in journals with significantly higher journal impact as compared to the lower performance universities. Moreover, the top universities perform in terms of citations per publications (***CPP***) with a factor of about 1.3 better than the bottom universities in journals with the same average impact. The relation between number of citations and field citation density found in this study can be considered as a second basic scaling rule of the science system.

Fifth, we find a significant decrease of the fraction of self-citations as a function of research performance ***CPP/FCSm***, of the average field citation density ***FCSm***, of the average journal impact ***JCSm***, and of the field-normalized journal impact ***JCSm/FCSm***.

***Acknowledgements***

*The author would like to thank his CWTS colleagues Henk Moed and Clara Calero for the work to define and to delineate the universities, and for the data collection, data analysis and calculation of the bibliometric indicators.*